\documentclass[usenatbib,onecolumn]{mn2e}
\usepackage{amsfonts}
\usepackage{amsmath}
\usepackage{amssymb}
\usepackage{graphicx}
\usepackage{natbib}
\usepackage{subfig}

\def\apjl{Astrophys.\ J.\ Lett.}
\def\mnras{Mon.\ Not.\ R.\ Astron.\ Soc.}

\def\apj{Astrophys.\ J.}

\def\prd{Phys.\ Rev.\ D}

\def\physrep{Phys. Rep.}
\def\pasj{Publ. Astron. Soc. Japan}

\title[Pairwise Velocity PDF in the $f_{nl}$ model]
      {The pairwise velocity probability density 
       function in models with local primordial non-Gaussianity}

\author[T. Y. Lam, T. Nishimichi \& N. Yoshida]
 {Tsz Yan Lam$^{1}$\thanks{E-mail:tszyan.lam@ipmu.jp},
  Takahiro Nishimichi$^{1}$\footnotemark[1] \& Naoki Yoshida$^1$\footnotemark[1] \\
$^1$ Institute for the Physics and Mathematics of the Universe, University of Tokyo, Kashiwa, Chiba 277-8583, Japan}

\newcommand{\bm}[1]{{\mbox{\boldmath $#1$}}}
\newcommand{\lrangle}[1]{\langle #1 \rangle}
\newcommand{\lpa}{_{\parallel}}
\newcommand{\lpe}[1][]{_{\perp_{#1}}}

\begin{document}
\pagerange{\pageref{firstpage}--\pageref{lastpage}}

\maketitle

\label{firstpage}

\begin{abstract}
We study how primordial non-Gaussianity affects the pairwise
velocity probability density function (PDF) using an analytical
model and cosmological $N$-body simulations. 
We adopt the local type non-Gaussian models characterized 
by $f_{nl}$, and examine both the linear velocity difference 
PDF
and the 
linear pairwise velocity PDF.
We show explicitly how $f_{nl}$ induces correlations between originally
independent velocities along the parallel and the perpendicular to the
line of separation directions.
We compare the model results with 
measurements from $N$-body simulations of the non-Gaussian models.
Linear theory fails to predict the PDF in the $f_{nl}$ models.
Therefore we develop an analytic model based on the Zeldovich approximation 
to describe the evolution of the velocity PDF. 
Our analytical model and simulation results 
show remarkably good agreement in both the parallel 
and the perpendicular directions for the PDF profiles, as well 
as the change in the PDF due to primordial non-Gaussianity.
The agreement is particularly good for relatively small separations 
($< 10 h^{-1}{\rm Mpc}$).
The inclusion of the evolution of the velocity PDF is
important to obtain a good description on the signature of primordial
non-Gaussianity in the PDF. 
Our model provides the foundation to constrain
$f_{nl}$ using the peculiar velocity  in future surveys.
\end{abstract}


\section{Introduction}
Primordial non-Gaussianity has attracted much attention owing to
its ability to distinguish inflationary models 
\citep[e.g.,][and references therein]{bko08,kp08,st08,wandslocalfnl}.
Several cosmological probes using the CMB 
\citep{cmb7yr,hikageetal08,yw08,mhlm08,rsphmfnl,cmbbktk,cmbfnlreview,
rossicmbfnl} and 
the large-scale structures in the Universe 
\citep{kst99,mvj00,ssz04,sk07,is07,fnlverde,dalaletal08,mv08,cvm08,
at08,shshp08,mcdonald08,tkm08,slosar08,grossi08,kvj08,fnlvincent,pphfnl08,
lamshethfnl,grossinfm,lsdfnl,lsdfnlred,nishimichifnl,CHDfnl,sbfmmrw10,LSSfnlreview,
cmblssfnlreview,ths10,nraofnl,chongchitnansilk10,gp10,laurafnlWL,robertfnlhaloModel} 
have been considered.

Here, we introduce a new large-scale structure probe of primordial
non-Gaussianity -- the pairwise velocity PDF. 
Current large-scale structure probes of primordial non-Gaussianity
focus on the clustering of halos, 
halo/void abundances, bispectrum, and the PDF of 
dark matter field -- all of them are related to the change of the density 
field due to primordial non-Gaussianity. 
It is important to note that both the initial density field and
the initial velocity field are generated from the primordial
perturbation, and that their linear relation is described by the continuity
equation. 
Hence, one may naively expect that primordial non-Gaussianity affect 
the velocity field as well as the density field.
While various measurements associated with the change of
density field have been extensively studied, there are few studies on the
effect of primordial non-Gaussianity on the velocity
field.
\citet{scherrer92,cs95} discussed the effect of primordial
non-Gaussianity on the distributions of linear velocity fields. In
particular \citet{cs95} studied the distribution of the parallel
component of the pairwise velocity using linear theory.
\citet{lsdfnlred} discussed the effect of primordial
non-Gaussianity on the redshift space distortion. They made
use of the ellipsoidal collapse model to derive the modification in
the Kaiser factor
relating the real and redshift space power spectra, without discussing
the modification in the velocity field.
During the preparation of this work, \citet{schmidt10} studied the primordial
non-Gaussianity signature in the peculiar velocities of density
peaks using linear theory. 
Our present study has two important improvements over these previous works: 
first we show that primordial non-Gaussianity
induces a correlation betweeen velocities in the parallel and the
perpendicular to the line of separation directions. 
We show explicitly how this correlation, which is absent in
\citet{schmidt10}, modifies significantly the linear 
velocity PDF.
Secondly, we show that the linear theory does not provide a good description 
of the signature of $f_{nl}$ in the pairwise velocity PDF even at
separation as big as 50 $h^{-1}{\rm Mpc}$. We use cosmological 
$N$-body simulations to show this.
We thus develop an analytic model to describe the evolution of the 
velocity PDF. Our model is based on the Zeldovich Approximation.
We illustrate the improvement in the $N$-body measurement comparisons.

The peculiar velocity field has been investigated as a probe of
cosmology for both dark matter
\citep{gorski88,seto98,shds01,kts02,rs04} 
and biased tracers 
\citep{sdvel01,sdhs01,hkys03,sz09}. 
Other studies \citep{ysd01,peel06,bk07,bk08sz} discuss the possibility
of constraining dark energy via
the kinetic Sunyaev-Zeldovich effect.
In this study we focus on the peculiar velocity field of the dark
matter field. The effect of primordial non-Gaussianity on the velocity
of biased tracers will be discussed in future work.

We first describe how the linear 
velocity PDF changes
due to  primordial non-Gaussianity 
in section~\ref{section:linearPDF}. 
Throughout the present paper, we work with models with
non-vanishing primordial bispectrum; we
will use the local $f_{nl}$ type primordial non-Gaussianity to
illustrate the calculations. 
The Bardeen potential $\Phi$ in the local $f_{nl}$ model is 
\begin{equation}
 \Phi = \phi + f_{nl}(\phi^2 - \langle \phi^2\rangle),
 \label{eqn:fnl}
\end{equation}
where $\phi$ is a Gaussian potential field and $f_{nl}$ is the
nonlinear quadratic parameter. 
The above  non-Gaussian
correction is defined at $z=z_{CMB}$ for this study.
It has been suggested that
mass weighting is important
for the peculiar velocities \citep{rs04,sz09}. 
We will discuss the effect of primordial non-Gaussianity on both the 
uniform weighted as well as the mass
weighted linear 
velocity PDF.
In section~\ref{section:evolvedPDF} 
we describe the analytical model to approximate the evolution of
the pairwise velocity PDF. 
The theoretical predictions of both the linear theory and the analytical
model are compared with measurements from $N$-body
simulations. We conclude our findings in section~\ref{section:discussion}.

\section{Linear velocity PDF} 
\label{section:linearPDF}

\subsection{Preliminary}
The linear overdensity $\delta(k)$ and the Bardeen potential $\Phi(k)$
is related by the Poisson equation
\begin{equation}
\delta(k,z) = D(z) k^2 M(k) \Phi(k),
\end{equation}
where  $M(k) = 2c^2T(k)/3\Omega_mH_0^2$ and 
$T(k)$ is the matter transfer function (note we do not include the
$k^2$ factor in $M(k)$). 
The continuity relates the linear overdensity and the peculiar
velocity $\bm{u}(\bm{k},z)$:
\begin{equation}
\dot{\delta}({\bm k}) + \theta({\bm k}) =0,
\end{equation}
where $\theta(\bm{x}) \equiv \nabla \cdot \bm{u}(\bm{x})$ is the
divergence of the velocity field. The velocity field is 
described solely by its divegerence since its vorticity decays due to the
expansion of the universe \citep[for example see][]{ptreview}. 
Hence
\begin{equation}
u_j( k) = i\dot{D}(z) k_jM(k) \Phi(k) , 
\end{equation}
where $i^2 = -1$ and the subscript $j$ denotes the coordinate of the
peculiar velocity. In this study we will be interested in the relative
velocities in the parallel ($\parallel$) and 
perpendicular ($\perp_a$ and $\perp_b$) to the line of separation for
two particles separated by some distance $r$.

Connected moments higher than the second order vanish when the
Bardeen potential is Gaussian. 
When the primordial perturbation is non-Gaussian,
the leading order of the non-vanishing
connected moment (higher than the second order) depends on the particular model of primordial
non-Gaussianity. 
Most studies in the literature concern with models with a non-vanishing bispectrum
(including the local $f_{nl}$ and the equilateral triangle type $f_{nl}$). 
Some other studies investigate models with a non-vanishing trispectrum 
\citep[the $g_{nl}$ model, see for example][]{vincentgnl}.
While this work focuses on primordial non-Gaussianity with a
leadingly non-vanishing bispectrum, it can be extended to 
study models
where the leading order of the non-vanishing connected moment 
is higher than the third
order. In what follows we will use the local $f_{nl}$ model to
demonstrate
how the non-vanishing third order connected moment modifies the
pairwise velocity PDF. The calculation can also apply to models with
other types of primordial bispectrum.

We denote the peculiar velocity at position $\bm x$ by
${\bm u}({\bm x})$ and the 
relative velocity by 
${\bm v}(\bm {r})\equiv {\bm u}({\bm x}) - {\bm u}({\bm x'}) = {\bm
  u}- {\bm u'} $ where ${\bm x} - {\bm x'} = {\bm r}$.
We also denote the velocity difference PDF $p({\bm v}; r)$ as the PDF 
of the peculiar velocity difference at two random
positions separated by $r$; 
the pairwise velocity PDF $q({\bm v}; r)$ as the PDF of the difference
of the peculiar velocities of two tested particles separated by
$r$. The latter PDF is the pair weighted version of the former one and
they are in general not equivalent \citep[see, for example][]{rs04}.
For simplicity we sometimes call the former PDF the 
uniform weighted PDF and the latter one the mass/pair weighted PDF.

The numerical integration of the third order connected moments are
carried out by Monte-Carlo integration using the numerical package
\textit{CUBA} \citep{cuba}.

\subsection{Linear Velocity Difference PDF}
\subsubsection{Case: $f_{nl} =0$}
The linear  velocity difference PDF of two random positions separated by a distance
$r$ when $f_{nl}=0$ is given by the
multivariate normal distribution:
\begin{equation}
p_0(\bm{v};r) = \frac{1}{(2\pi)^{3/2}\sqrt{|A|}}\exp\left(-\frac{1}{2}\bm{v}^TA^{-1}\bm{v}\right),
\end{equation}
where $\bm{v} = (v\lpa ,v\lpe[a] ,v\lpe[b] )$ in which
$v\lpa $ corresponds to the 
relative velocity parallel to the line
of separation and $v\lpe[a] $ and $v\lpe[b] $ are the two 
velocity differences perpendicular to the line of separation. 
Here $A$ is the
covariance matrix. The above expression simplifies since there is no correlation
between different components of $\bm{v}$. Hence the r.h.s becomes a
product of three univariate normal distributions:
\begin{equation}
p_0(v\lpa ,v\lpe[a] ,v\lpe[b] ; r) =
p_0(v\lpa ;r)p_0(v\lpe[a] ;r) p_0(v\lpe[b] ;r),
\label{eqn:pdfgvw}
\end{equation}
where the variances of the univariate normal distributions are 
$\langle v\lpa ^2 \rangle$ and $\langle \mathfrak{v}_{\perp}^2 \rangle
\equiv \langle v\lpe[a] ^2 \rangle= \langle v\lpe[b] ^2 \rangle$
respectively, and
\begin{align}
\langle v\lpa ^2 \rangle & = \frac{1}{3\pi^2}\dot{D}^2_0\int {\rm
  d} k\,P_{\Phi}(k) k^4M^2(k)\left[1- 3j_0(kr) +
  6\frac{j_1(kr)}{kr}\right] \\
\langle \mathfrak{v}_{\perp}^2 \rangle & = \frac{1}{3\pi^2}\dot{D}^2_0\int{\rm d}
k\, P_{\Phi}(k)k^4M^2(k)\left[1 - 3\frac{j_1(kr)}{kr}\right].
\end{align}
Here $j_0$ and $j_1$ is the spherical bessel function, $\dot{D}_0 =
{\rm d} D/ {\rm d} t$ is the time derivative of the linear growth
factor $D$    at $z=0$ and $P_{\Phi}(k) \approx P_{\phi}(k)$ is the
Bardeen potential power spectrum. 
Iostropy means one can transform the rectangular coordinate in the plane
perpendicular to the line of separation into the polar coordinate and
write equation~\eqref{eqn:pdfgvw} as
\begin{equation}
p_0(v\lpa ,v\lpe[a] ,v\lpe[b] ; r)dv\lpa dv\lpe[a] dv\lpe[b] = 2\pi
v\lpe p_0(v\lpa;r)p_0(v\lpe ; r) dv\lpa dv\lpe , 
\end{equation}
where $v\lpe ^2= v\lpe[a]^2 + v\lpe[b]^2 $. 
Notice that our notations 
$\lrangle{\mathfrak{v}\lpe\cdots} (= \lrangle{v\lpe[a]\cdots}
=\lrangle{v\lpe[b]\cdots})$ and $v\lpe (= \sqrt{v\lpe[a]^2 + v\lpe[b]^2}) $
are not equivalent.

\subsubsection{Case: $f_{nl} \neq 0$}
When $f_{nl}\neq 0$, the primordial bispectrum is
non-zero and its functional form  
in the local type is 
\begin{equation}
B_{\Phi}(k_1,k_2,k_{12}) = 2f_{nl} [P(k_1)P(k_2) + {\rm cyclic}] + \mathcal{O}(f_{nl}^3). 
\end{equation}
Connected moments higher than the second
order are non-vanishing and contribute to the  velocity difference PDF.
The leadingly non-vanishing connected moments are 
\begin{align}
\frac{\langle v\lpa ^3 \rangle}{2f_{nl}} & =
\frac{12\dot{D}^3_0}{(2\pi)^6}\int{\rm d}^3{\bm k_1}\int_{\cos\mu_2\geq 0}{\rm d}^3{\bm k_2}
P(k_1)P(k_2)M(k_1)M(k_2)M(k_{12}) k_{1\parallel} k_{2\parallel}
k_{12\parallel} [\sin(k_{12\parallel}r) - 2\sin(k_{2\parallel}r)]   \\
\frac{\langle v\lpa \mathfrak{v}_{\perp}^2\rangle}{2f_{nl}} & \equiv \frac{\langle
v\lpa v\lpe[a] ^2\rangle}{2f_{nl}} = \frac{\langle
v\lpa v\lpe[b] ^2\rangle}{2f_{nl}}    
 = \frac{4\dot{D}^3_0}{(2\pi)^6}\int{\rm d}^3{\bm
  k_1}\int_{\cos\mu_2\geq 0}{\rm d}^3{\bm
  k_2}P(k_1)P(k_2)M(k_1)M(k_2)M(k_{12}) [k_{1\perp}k_{2\perp}k_{12\parallel}\sin(k_{12\parallel}r)  \nonumber \\
& \qquad    -2k_{1\perp}k_{2\parallel}k_{12\perp}\sin(k_{2\parallel}r) 
    + 2k_{1\perp}k_{2\parallel}k_{12\perp}\sin(k_{12\parallel}r)
     -2k_{1\parallel}k_{2\perp}k_{12\perp}\sin(k_{2\parallel}r)
     - 2k_{1\perp}k_{2\perp}k_{12\parallel} \sin(k_{2\parallel}r)],
\end{align}
where $k_{12} = |k_1 + k_2|$, $k_{i\parallel} = k_i\cos\mu_i$,
$k_{i\perp} = k_i\sin\mu_i\cos\phi_i$, and $k_{12\parallel} =
k_{1\parallel} + k_{2\parallel}$ (similarly for $_{\perp}$).
Note that connected moments involving odd power of $v\lpe$ vanish
irrespective of the value of $f_{nl}$. 
The non-vanishing component $\langle v\lpa \mathfrak{v}_{\perp}^2\rangle$
results in couplings between velocities in the parallel and the
perpendicular to the line of separation directions. As a result the linear
multivariate PDF can no longer be written as  
a product of three independent univariate PDFs.
We use the tri-variate Edgeworth expansion to approximate the linear
uniform weighted velocity difference PDF 
for $f_{nl} \neq 0$ (see appendix for derivation):
\begin{equation}
p(v\lpa ,v\lpe[a] ,v\lpe[b] ;f_{nl},r) =
p_0(v\lpa ,v\lpe[a] ,v\lpe[b] ;r)
[1 + \alpha_{300} h_{300} + \alpha_{120}(h_{120} + h_{102})],
\label{eqn:pdffnlvw}
\end{equation}
where 
\begin{equation*}
\alpha_{300}  = \frac{1}{6}\frac{\langle v\lpa ^3\rangle}{\langle
   v\lpa ^2\rangle^{3/2}}, \qquad
\alpha_{120}  = \frac{1}{2}\frac{\langle v\lpa \mathfrak{v}_{\perp}^2\rangle}{\langle
  v\lpa ^2\rangle^{1/2} \langle \mathfrak{v}_{\perp}^2\rangle}, 
\end{equation*}
and 
$h_{ijk} \equiv H_i(\nu\lpa )H_j(\nu\lpe[a] )H_k(\nu\lpe[b] )$ is the
product of Hermite polynomials of different orders. 
In particular, 
\begin{align}
h_{300} & \equiv \nu\lpa ^3 - 3\nu\lpa  = 
\frac{v\lpa ^3}{\langle v\lpa ^2\rangle^{3/2}} - 3
\frac{v\lpa }{\langle v\lpa ^2\rangle^{1/2}} \\
h_{120} & \equiv  \nu\lpa   (\nu\lpe[a] ^2 - 1) = 
\frac{v\lpa }{\langle v\lpa ^2\rangle^{1/2}}\left(\frac{
    v\lpe[a] ^2}{\langle \mathfrak{v}_{\perp}^2\rangle } -1 \right) \\
h_{102} & \equiv  \nu\lpa   (\nu\lpe[b] ^2 - 1) = 
\frac{v\lpa }{\langle v\lpa ^2\rangle^{1/2}}\left(\frac{
    v\lpe[b] ^2}{\langle \mathfrak{v}_{\perp}^2\rangle } -1 \right).
\end{align}
Note that \citet{schmidt10} 
marginalized velocities perpendicular to the line of separation,
essentially setting $\alpha_{120}=0$.

\begin{figure}
\centering
\includegraphics[width=0.7\textwidth]{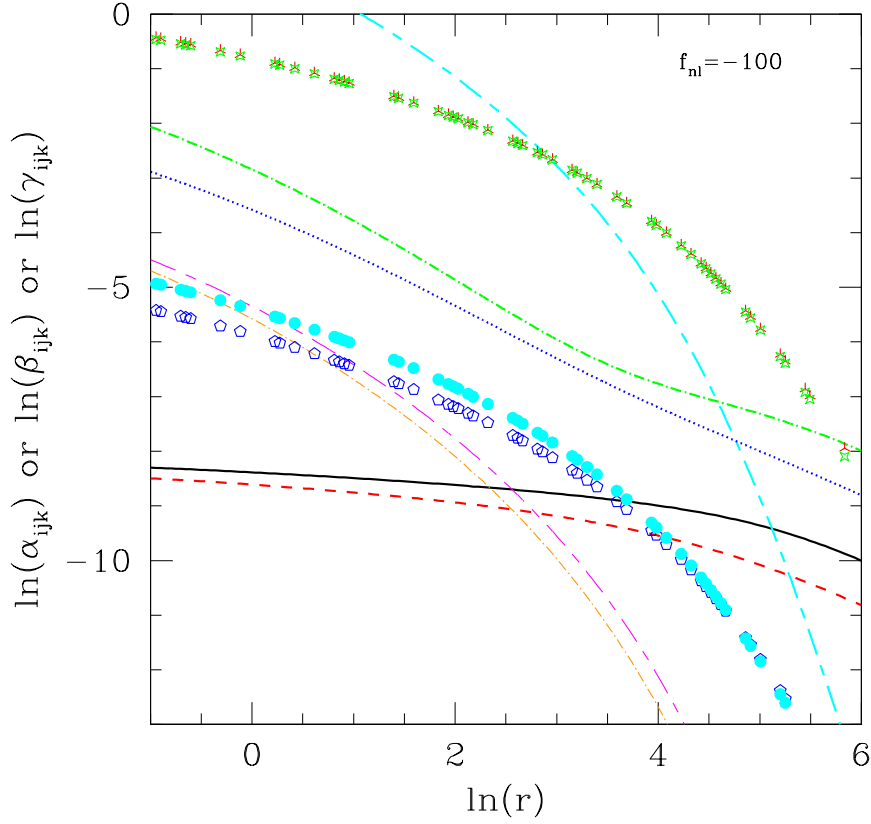}
\caption{Scale dependence of parameters: $\alpha_{300}$ (thick solid black
  curve), $\alpha_{120}$ (thick dashed red curve),$-\beta_{100}$ (red
  sketeal trianglar symbols), $\beta_{200}$ (thick cyan
  short-long-dashed curve), $-\gamma_{100}$ (green starred
  symbols), 
  $\gamma_{300}$ (thick green dot-dashed 
  curve), $\gamma_{120}$ (thick blue dotted curve), $-\gamma_{400}$
  (blue pentagons), $-\gamma_{220}$ (cyan hexagons), $\gamma_{500}$
  (thin magenta short-long-dashed curve), and $\gamma_{320}$ thin (orange
  dot-dashed curve). $f_{nl} = -100 $ for those parameters
  apply. $\alpha_{300}$, $\alpha_{120}$, $-\gamma_{100}$, and
  $-\beta_{100}$ are shifted vertically by -2 for clarity.}
\label{fig:paraVW}
\end{figure}
Figure~\ref{fig:paraVW} shows the scale dependence of $\alpha_{300}$
(thick solid black curve) and $\alpha_{120}$ (thick dashed red curve)
for $f_{nl} = -100$.
While the numerical value is small compared to unity, the
effect of non-zero $f_{nl}$ becomes significant  
for big $|\nu\lpa|$ or $|\nu\lpe|$.
Furthermore  $\alpha_{300}$ and $\alpha_{120}$ are
in the same order of magnitude, 
hence both terms have to be included in the computation of 
the linear peculiar velocity difference PDF.

\begin{figure}
\centering
\includegraphics[angle=-90,width=0.7\textwidth]{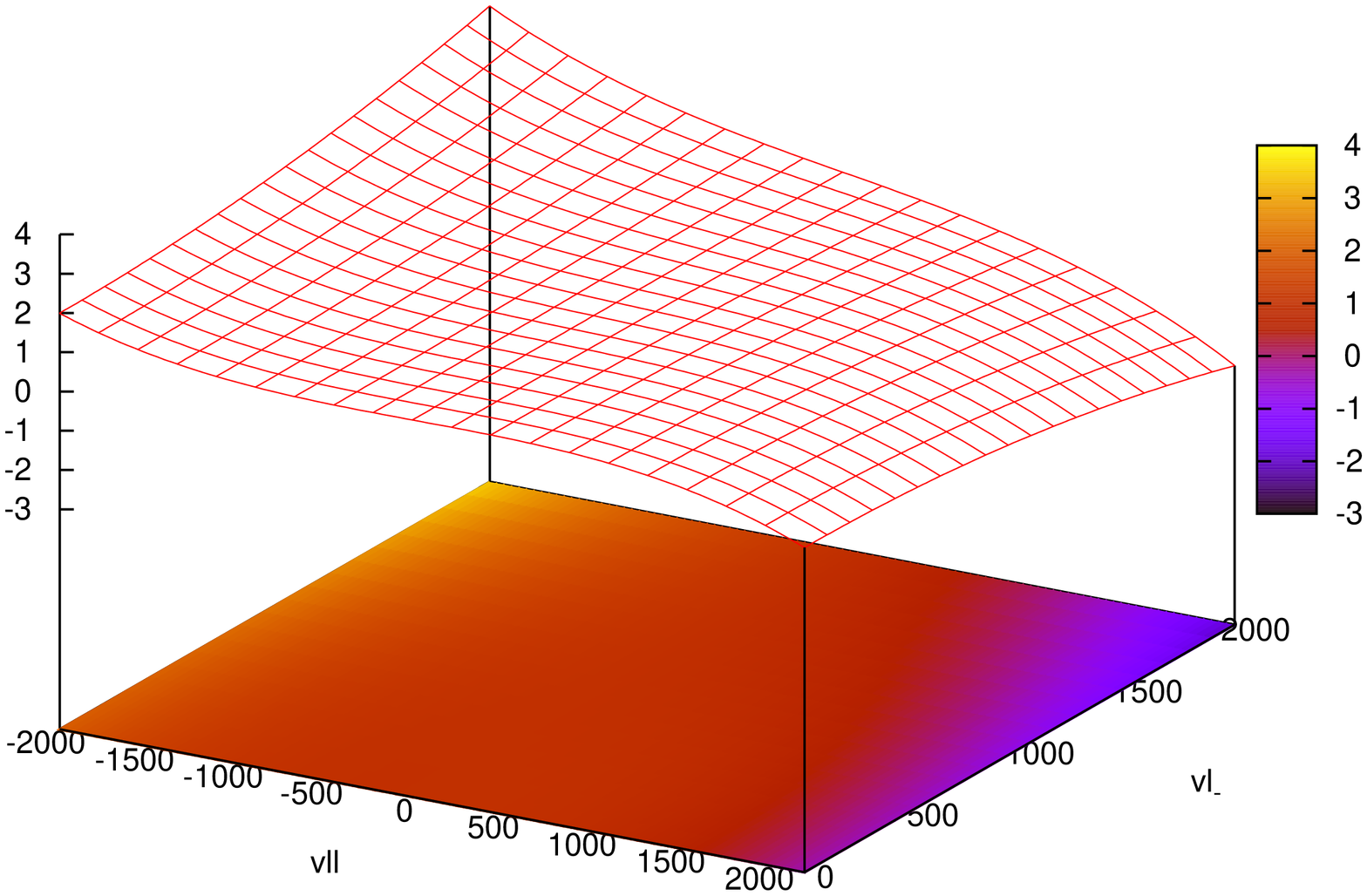}%
\vspace{-10pt}
   \begin{minipage}[c]{0.5\linewidth}
     \centering %
     \includegraphics[angle=-90,width=\textwidth]{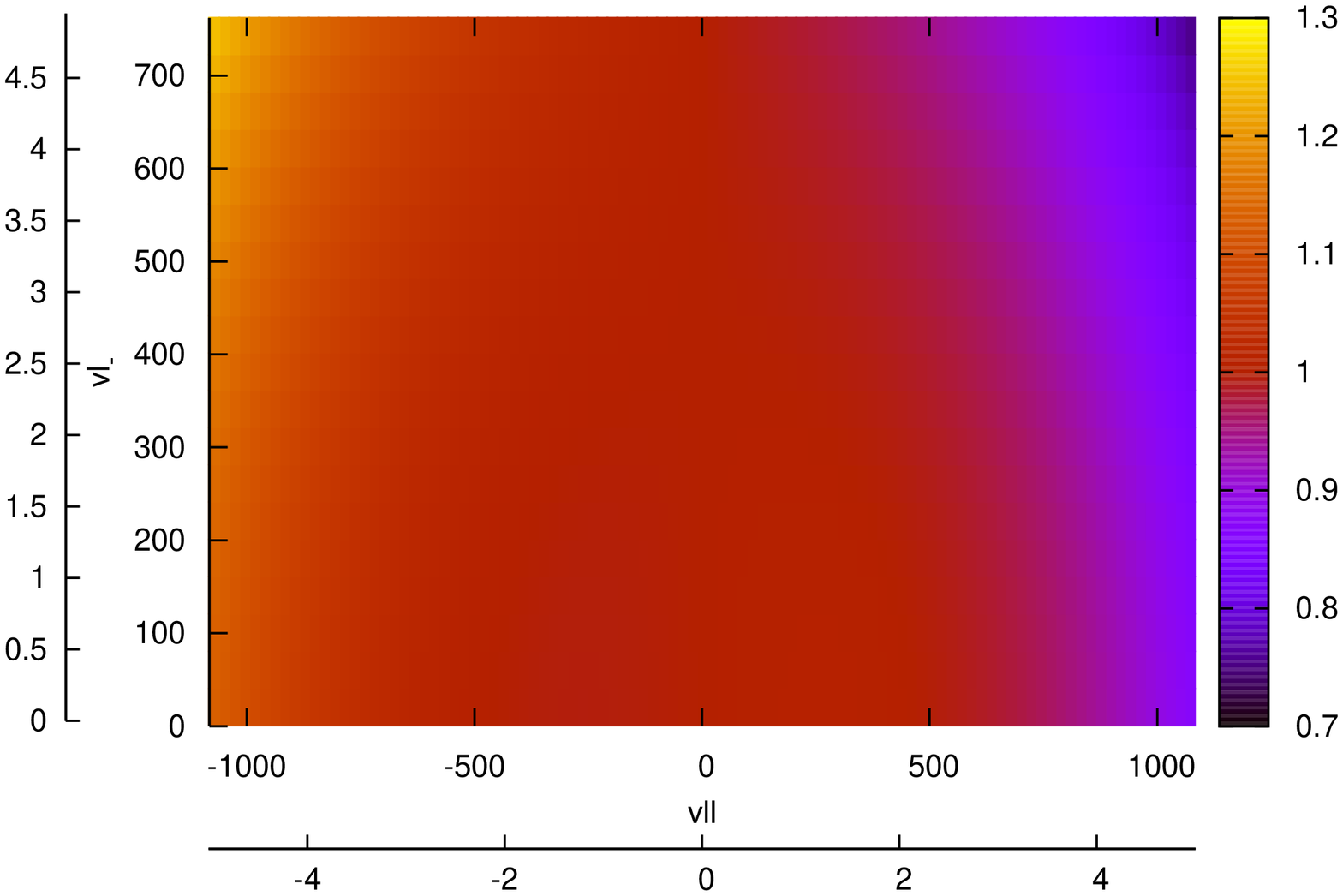}
   \end{minipage}%
   \begin{minipage}[c]{0.5\linewidth}
      \centering %
      \includegraphics[angle=-90,width=\textwidth]{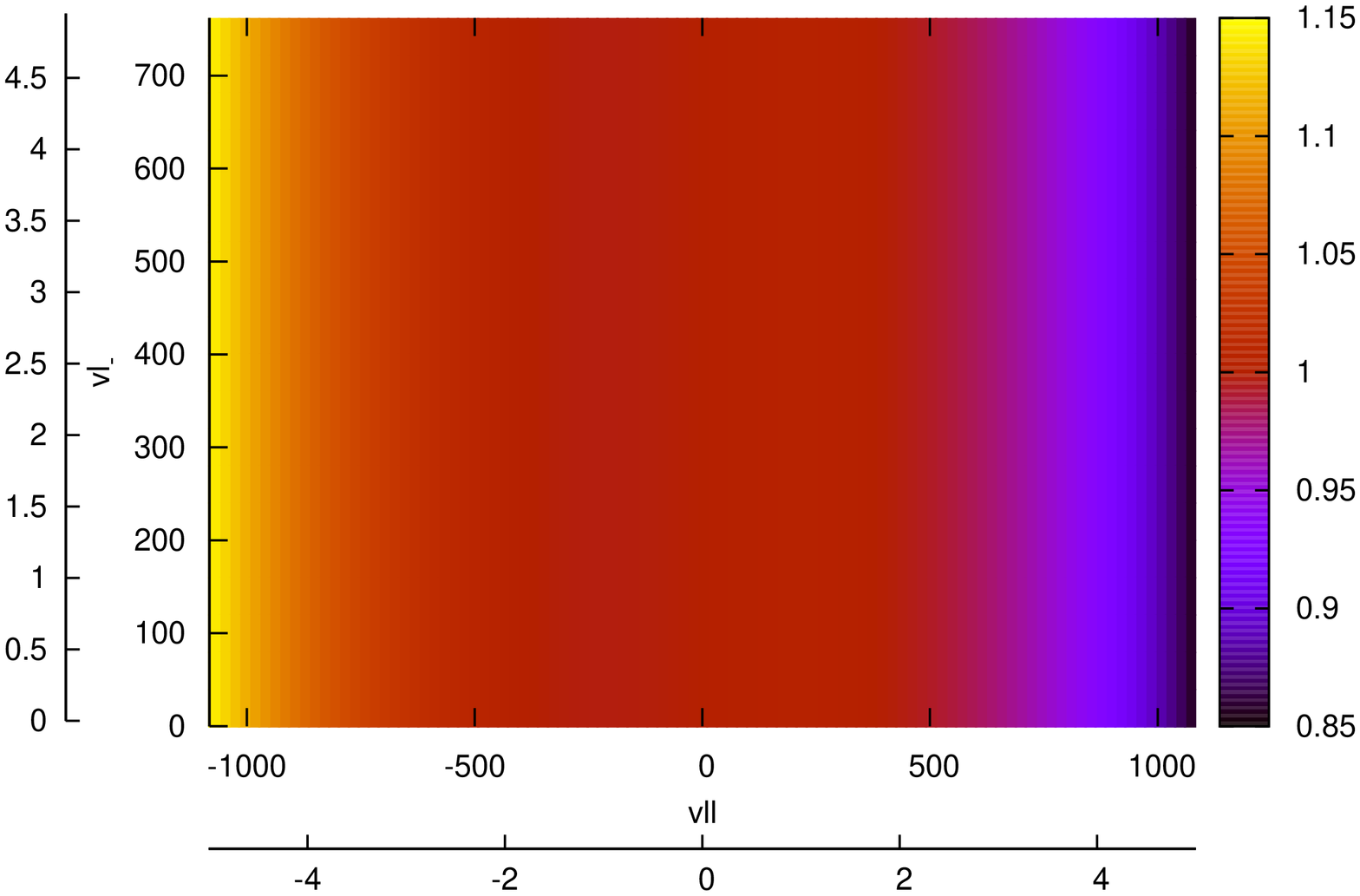}
    \end{minipage}  
\vspace{2.cm}
\caption{Ratio of linear velocity difference PDF $p/p_0$ in
  equation~\eqref{eqn:pdffnlvw} for $r=8\ h^{-1}{\rm Mpc}$ and $f_{nl} =
  100$. vll indicates velocity in the parallel to the line of
separation direction while ${\rm vl\_}$ is the magnitude of the velocity 
perpendicular to the line of separation. 
 Upper panel shows the ratio for a wide range of velocity. The 
 lower left panel zooms into regions for around 5-$\sigma$ as indicated by
 the outer axes. The lower right panel shows the same regions as the
 lower left panel, but explicitly set the parameter $\alpha_{120} =0$
 as previous study did.}
\label{fig:pdfvlinearVW}
\end{figure}
Figure~\ref{fig:pdfvlinearVW} shows the ratios of the linear peculiar
velocity difference PDF for $f_{nl}=
100$ to the corresponding Gaussian PDF at 
$r = 8\ h^{-1}{\rm Mpc}$. 
The axes labeled vll indicates the velocity in the parallel to the line of
separation direction and ${\rm vl\_}$ is the magnitude of the velocity 
perpendicular to the line of separation (recall ${\rm vl\_} =
\sqrt{v\lpe[a]^2 + v\lpe[b]^2}$).
The upper panel
shows both the contour and color maps over a wide range of velocity. 
The effect of primordial non-Gaussianity is most significant at 
extreme velocities. 
Both the contour and the color maps show 
variations in the ratio for different values of
$v_{\perp}$ at fixed $v\lpa $, indicating that 
the modification due to $f_{nl}$ is degenerated in the plane of 
${\rm (vll,vl\_)}$.
While the upper panel shows that the ratio can be 
as big as 4, 
the first order Edgeworth
expansion breaks down in such rare cases (about 10-$\sigma$ level) --
it is evident by the negative probability at the
other end of the plot.

The lower panels of Figure~\ref{fig:pdfvlinearVW} show the color maps
of the same PDF ratio on a smaller velocity range to increase the
dynamic range of 
the effect of $f_{nl}$ on the linear velocity difference PDF. 
The left panel shows the modification obtained in this study (the
correction term in equation~\eqref{eqn:pdffnlvw}) while the right
panel shows the result for setting $\lrangle{v\lpa \mathfrak{v}\lpe^2}=0$.
The outer axes of the lower panels show the $\sigma$-level in corresponding 
directions. At $5$-$\sigma$ level (${\rm vll}$ direction) the change
is 
about $30\%$, 
and the bigger the $\sigma$-level in the perpendicular direction the
greater the modification is. 
The linear theory suggests that for $f_{nl}>0$ the infalling probability
(negative velocity) is enhanced while the outgoing probability
(positive velocity) is decreased. 
This trend applies to other scales we looked
at (from $4 $ to $100\ h^{-1}{\rm Mpc}$).
While it is generally believed that 
linear theory applies for large
scales, we will show in the next section that 
linear theory fails to describe 
the change in the
velocity PDF due to $f_{nl}$ even at separation as big as 50
$h^{-1}{\rm Mpc}$.

\subsection{Linear Pairwise Velocity PDF}
It is widely accepted that, while the linear theory prediction is a good
approximation for large scales ($\gtrsim 20\ h^{-1}{\rm Mpc}$), the
linear theory velocity correlations of massive halos (in the parallel to the line of
separation direction) is not consistent with $N$-body measurements \citep{ce94}. 
\citet{sz09} pointed out that the velocity correlations of biased
tracers, in both
parallel and perpendicular to the line of separation directions, 
can in fact be reasonably described by linear theory when
proper pair-weighting is included.
While our current study focuses on the difference of the 
peculiar velocities of unbiased tracer, 
it is useful to examine whether the pair weighting  would improve
the linear theory prediction.
It is reasonable to include this mass weighting as 
the pairwise velocity PDF measured in $N$-body simulations is obtained by
counting
the relative velocities of simulated particles. 
As a result regions with more particles (overdense
regions) have a bigger weight than regions with less particles
(underdense regions).

\subsubsection{Case: $f_{nl} = 0$}
\citet{rs04} discussed how to generalize the linear
 velocity difference PDF when $f_{nl}=0$ to include the mass weighting. 
The resulting linear  pairwise velocity PDF $q_0$ is
related to the associated no weighting velocity difference  PDF $p_0$ by
\begin{equation}
[1+\xi (r)]
\frac{q_0(\nu\lpa ,\nu\lpe[a] ,\nu\lpe[b] ;r)}{p_0(\nu\lpa ,\nu\lpe[a] ,\nu\lpe[b] ;r)} =
1 + \xi(r) + h_{100}\beta_{100} + h_{200}\beta_{200},
\label{eqn:q0p0}
\end{equation}
where  
$\delta \equiv \delta({\bm
  x})$,  $\delta' \equiv \delta({\bm x +r})$, 
$\xi\equiv \xi(r) = \langle\delta\delta'\rangle$ is the matter correlation
function at separation $r$, and
\begin{align}
\beta_{100} & = \frac{\langle v\lpa \delta\rangle}{\langle v\lpa ^2 \rangle^{1/2}} +
    \frac{\langle v\lpa \delta'\rangle}{\langle
        v\lpa ^2 \rangle^{1/2}},  \qquad
\beta_{200} = \frac{\langle v\lpa \delta\rangle \langle
    v\lpa \delta'\rangle}{\langle v\lpa ^2\rangle}, \\
\langle v\lpa \delta\rangle & = \langle
v\lpa \delta'\rangle = \frac{1}{2\pi^2}\dot{D}_0 \int {\rm d} k
P_{\Phi}(k)k^4 M^2(k) [kj_1(kr)]. 
\end{align} 
The r.h.s of equation~\eqref{eqn:q0p0} does not depend $v_{\perp}$,
hence $q_0$ can still be written as a product of three univariate PDFs --
the correction term on the r.h.s of equation~\eqref{eqn:q0p0} 
only modifies
the univariate PDF of $v\lpa $.
The scale dependence of the parameters $-\beta_{100}$ (red sketeal
triangular symbols) and $\beta_{200}$ (thick cyan short-long-dashed curve)
are shown in Figure~\ref{fig:paraVW}: their magnitudes
are big compared to 
others shown in the same figure since they
are 'Gaussian parameters' that do not depend on $f_{nl}$.

\subsubsection{Case: $f_{nl}\neq 0$}
When $f_{nl}\neq 0$ additional terms contribute to the mass weighted linear
 velocity difference PDF.
 To the first order of $f_{nl}$ the expression is (see derivation
in appendix)
\begin{align}
[1 + \xi(r)]  \frac{q(\nu\lpa ,\nu\lpe[a] ,\nu\lpe[b] ;
f_{nl},r)}{p_0(\nu\lpa ,\nu\lpe[a] ,\nu\lpe[b] ;r)}  & =
1 + \xi(r) + h_{100}\gamma_{100} + h_{200}\beta_{200}
+ h_{300}\gamma_{300} + (h_{120} + h_{102})\gamma_{120}  \nonumber \\
& \quad +
h_{400}\gamma_{400} + (h_{220} + h_{202})\gamma_{220}
+ h_{500}\gamma_{500} + (h_{320} + h_{302})\gamma_{320},
\label{eqn:qfnl}
\end{align}
where
\begin{align*}
\gamma_{100} & = \beta_{100}  +  \frac{\langle
  v\lpa \delta\delta'\rangle}{\langle
  v\lpa ^2\rangle^{1/2}} & \qquad
\gamma_{300} & = \alpha_{300} (1 + \xi(r)) + \frac{\beta_{100}}{2}\frac{\langle
   v\lpa ^2\delta'\rangle}{\langle
   v\lpa ^2\rangle} & \qquad
\gamma_{120} & = \alpha_{120}(1+\xi(r)) + \frac{\beta_{100}}{2}\frac{
     \langle \mathfrak{v}_{\perp}^2\delta'\rangle}{\langle \mathfrak{v}_{\perp}^2\rangle} \\
\gamma_{400} & = \alpha_{300}\frac{\beta_{100}}{2} & \qquad
\gamma_{220} & = \alpha_{120}\beta_{100} & \qquad
\gamma_{500} & = \alpha_{300}\beta_{200}\\
\gamma_{320} & = \alpha_{120}\beta_{200} & \qquad
h_{400} & = \nu\lpa ^4 - 6\nu\lpa ^2 +3 & \qquad
h_{500} & = \nu\lpa ^5 - 10 \nu\lpa ^3 +
15\nu\lpa , 
\end{align*}
and for the local $f_{nl}$ type primordial non-Gaussianity
\begin{align}
\frac{\langle v\lpa \delta\delta'\rangle}{2f_{nl}}
& =-\frac{4\dot{D}_0}{(2\pi)^6} 
\int{\rm d}^3{\bm k_1}\int_{\cos\mu_2\geq 0}{\rm d}^3{\bm k_2}
P(k_1)P(k_2)M(k_1)M(k_2)M(k_{12})
[k_{1\parallel}k_2^2k_{12}^2\sin(k_{12\parallel}r)  + k_1^2k_2^2
k_{12\parallel}\sin(k_{2\parallel}r)   \nonumber \\
& \qquad - k_{1\parallel}k_2^2k_{12}^2\sin(k_{2\parallel}r)] \\
\frac{\langle v\lpa ^2\delta'\rangle}{2f_{nl}} & =
\frac{2\dot{D}_0^2}{(2\pi)^6} 
\int{\rm d}^3{\bm k_1}\int_{\cos\mu_2\geq 0}{\rm d}^3{\bm k_2}
P(k_1)P(k_2)M(k_1)M(k_2)M(k_{12}) 
[2k_{1\parallel}k_2^2k_{12\parallel}\cos(k_{2\parallel}r) -
k_{1\parallel}k_{2\parallel}k_{12}^2\cos(k_{12\parallel}r) \nonumber \\
& \qquad + 2k_{1\parallel}k_{2\parallel}k_{12}^2\cos(k_{2\parallel}r) 
- 2k_{1\parallel}k_2^2k_{12\parallel}\cos(k_{1\parallel}r)  
  - 2k_{1\parallel}k_2^2k_{12\parallel}\cos(k_{12\parallel}r)
 + 2k_{1\parallel}k_2^2k_{12\parallel} -
 k_{1\parallel}k_{2\parallel}k_{12}^2] \\
\frac{\langle \mathfrak{v}_{\perp}^2\delta'\rangle}{2f_{nl}} & =
\frac{2\dot{D}_0^2}{(2\pi)^6} 
\int{\rm d}^3{\bm k_1}\int_{\cos\mu_2\geq 0}{\rm d}^3{\bm k_2}
P(k_1)P(k_2)M(k_1)M(k_2)M(k_{12}) 
[2k_{1\perp}k_2^2k_{12\perp}\cos(k_{2\parallel}r) -
k_{1\perp}k_{2\perp}k_{12}^2\cos(k_{12\parallel}r) \nonumber \\
& \qquad + 2k_{1\perp}k_{2\perp}k_{12}^2\cos(k_{2\parallel}r) 
- 2k_{1\perp}k_2^2k_{12\perp}\cos(k_{1\parallel}r)  
  - 2k_{1\perp}k_2^2k_{12\perp}\cos(k_{12\parallel}r)
 + 2k_{1\parallel}k_2^2k_{12\parallel} - k_{1\parallel}k_{2\parallel}k_{12}^2].
\end{align}
The scale dependences of $\gamma_{ijk}$ are shown in figure~\ref{fig:paraVW}.
As in the case of uniform weighted linear PDF, the linear pairwise
velocity PDF can no longer be written as a product
of three independent PDFs when $f_{nl} \neq 0$ due to the non-vanishing 
terms $\langle v\lpa \mathfrak{v}_{\perp}^2 \rangle$ and $\langle
\mathfrak{v}_{\perp}^2\delta'\rangle$.
In addition, when one sets $\delta = \delta' = 0$, the above expression
recovers the uniform weighted linear PDF 
for $f_{nl}\neq 0$ (equation~\eqref{eqn:pdffnlvw}).

\begin{figure}
\centering
   \includegraphics[angle=-90,width=0.7\textwidth]{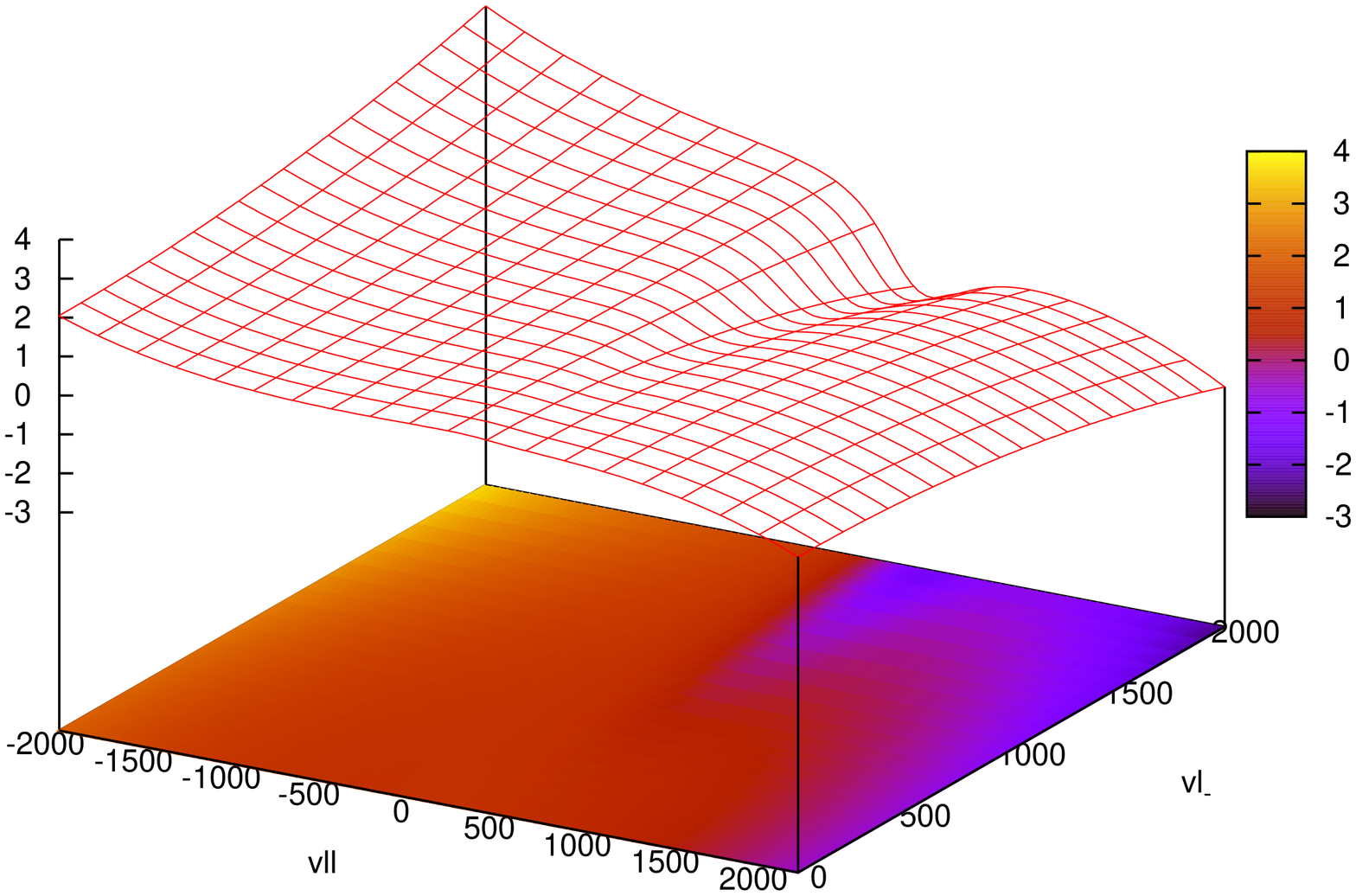}%
  \vspace{-10pt}
   \begin{minipage}[c]{0.5\linewidth}
     \centering %
     \includegraphics[angle=-90,width=\textwidth]{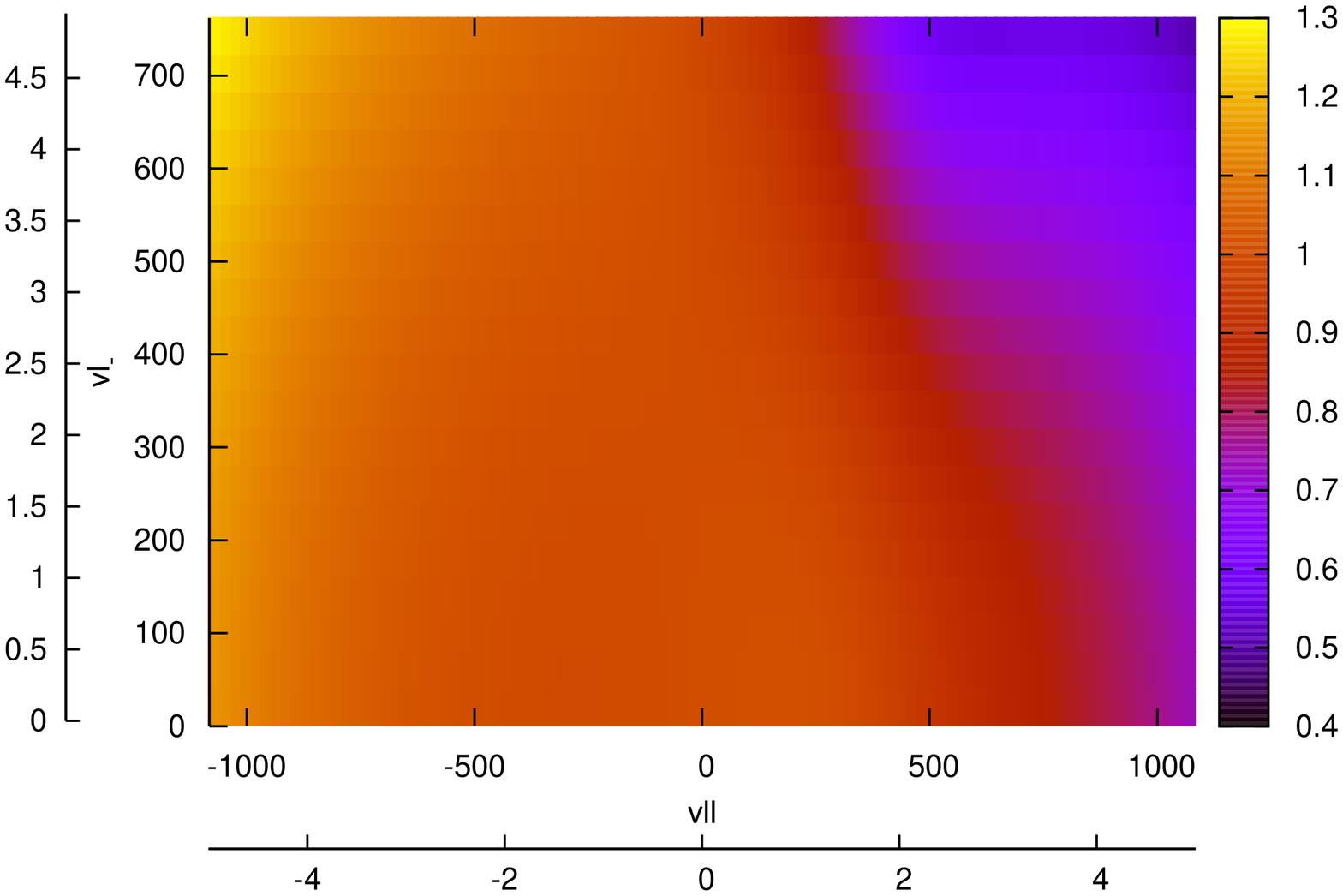}
   \end{minipage}%
   \begin{minipage}[c]{0.5\linewidth}
      \centering %
      \includegraphics[angle=-90,width=\textwidth]{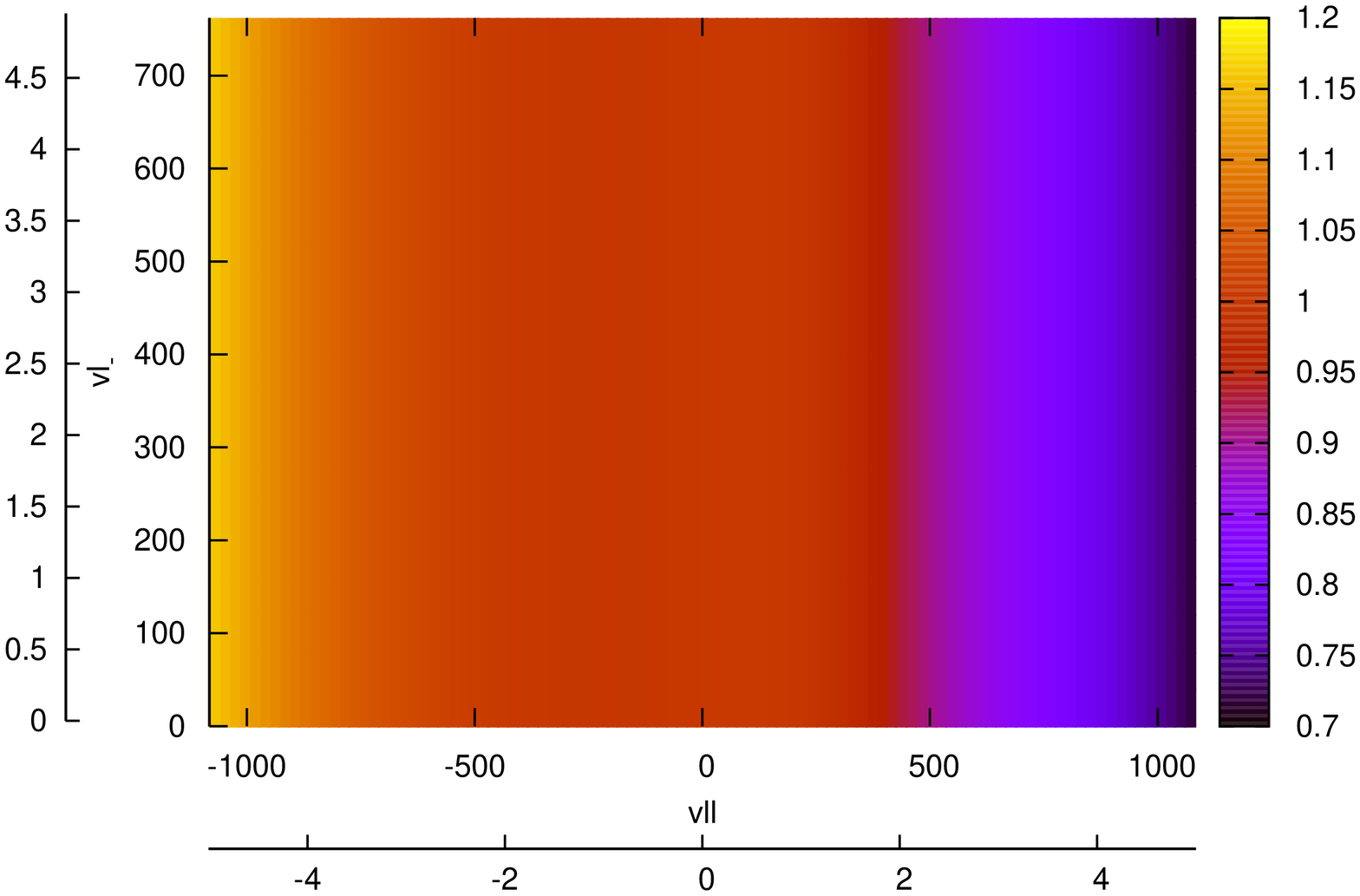}
    \end{minipage}  
\vspace{2.cm}
\caption{Similar plot to figure~\ref{fig:pdfvlinearVW} but for the
  ratio of the 
  linear pairwise velocity PDF ($q/q_0$ in
  equations~\eqref{eqn:q0p0} and \eqref{eqn:qfnl}). 
  The lower right panel
  sets  $\alpha_{120} = \gamma_{220} = \gamma_{320} 
            = \lrangle{\mathfrak{v}\lpe^2\delta'} = 0$.}
\label{fig:pdfvlinearMW}
\end{figure}
Figure~\ref{fig:pdfvlinearMW} shows the modification factor of the 
 linear pairwise velocity PDF due to $f_{nl}$
($q/q_0$ in equations~\eqref{eqn:q0p0} and \eqref{eqn:qfnl}). 
As in Figure~\ref{fig:pdfvlinearVW} the upper panel shows the ratio
in a wide range of velocity, including regions with negative
probability indicating the breakdown of the first order Edgeworth
expansion approximation. 
The lower left panel enlarges the color map on regions about
5-$\sigma$ level
and 
the outer axes label the $\sigma$-level.
The lower right panel shows the ratio if one explicitly sets all third
order moments involving 
$v\lpe$
to zero 
(i.e. 
$\alpha_{120} = \gamma_{220} = \gamma_{320} = \lrangle{\mathfrak{v}\lpe^2\delta'}
= 0$).
The effect of $f_{nl}$ in the mass weighted linear pairwise PDF has the
same trend as the uniform weighted PDF:
the effect of $f_{nl}$ is degenerated in the parallel
and the perpendicular to the line of separation directions;
and the general trend observed in uniform weighted
linear PDF still applies -- for $f_{nl} > 0$
the probability of finding infalling pairs increases while the
probability of finding pairs moving apart decreases.
The inclusion of the mass weighted quantites $\gamma_{ijk}$ strengthens  the
effect of $f_{nl}$ and it is most significant in the decrement in probability
of outgoing pairs.

\section{Evolved velocity PDF}
\label{section:evolvedPDF}
Given the linear velocity PDF we now describe a model for the
evolution of the PDF. 
We adopt the Zeldovich
Approximation which assumes the comoving 
velocity remains unchanged.
At some initial redshift $z_i$, if two particles separated by $r_i$
have a relative velocity
$(v\lpa ^i,v\lpe[a] ^i,v\lpe[b] ^i)$. 
Then at a later redshift $z=z_0 \ (\ll z_i)$ the
distances traveled 
are 
$D_0/\dot{D}_i (v\lpa ^i,v\lpe[a] ^i,v\lpe[b] ^i)$ 
(note that it assumes 
$D_0 \gg D_i$).
Hence the separation of the two particles becomes
\begin{equation}
r^2 = \left( r_i + \frac{D_0}{\dot{D}_i} v\lpa ^i\right)^2 +
\left(\frac{D_0}{\dot{D}_i}\right)^2({v\lpe[a] ^i}^2 +
{v\lpe[b] ^i}^2).
\label{eqn:rsq}
\end{equation}
The evolved 
relative velocities (with respect to the updated position)
 are:
\begin{align}
v\lpa  & = \frac{\dot{D}_0}{r}\left(\frac{r_i
    v\lpa ^i}{\dot{D}_i} + \frac{D_0}{\dot{D}_i^2}
  {v^i}^2\right) \label{eqn:vll} \\
|v_{\perp}|^2 & = v\lpe[a] ^2  + v\lpe[b] ^2 =
\left(\frac{\dot{D}_0}{\dot{D}_i}{v^i}\right)^2 - v\lpa ^2 \label{eqn:vl_},
\end{align}
where ${v^i}^2 = {v\lpa ^i}^2 + {v\lpe[a] ^i}^2 + {v\lpe[b] ^i}^2$.
The evolved velocity difference PDF is therefore 
\begin{equation}
p(V\lpa ,V_{\perp};R) = \int {\rm d}r_i {\rm
  d}v\lpa ^i{\rm d} v\lpe[a] ^i{\rm d} v\lpe[b] ^i
\frac{r_i^2}{R^2} p(v\lpa ^i,v\lpe[a] ^i,v\lpe[b] ^i;r)
\delta_{\rm D}(r - R)\delta_{\rm D}(v\lpa  - V\lpa )
\delta_{\rm D}(v_{\perp} - V_{\perp}), 
\label{eqn:evolvedpdf}
\end{equation}
\citep[see, for example,][]{seto98}
where the dirac-delta functions use equations~\eqref{eqn:rsq},
\eqref{eqn:vll}, and \eqref{eqn:vl_} to map the initial quantities to
the evolved one. 
In the following we apply the above formula 
to compute the evolved velocity difference PDF from 
equations~\eqref{eqn:pdfgvw} and \eqref{eqn:pdffnlvw}
and
compare the results to measurements from $N$-body simulations.
We set $z_i$ and $z_0$ to be $1100$ and $0.5$ respectively, justifying 
the approximation $D_0 \gg D_i$.

\subsection{Comparisons to $N$-body measurements}
We measured the pairwise velocity PDF from a set of $N$-body
simulations described in \citet{nishimichifnl}. 
The simulation adopts the \emph{WMAP} 5-years $\Lambda$CDM best fit
parameters $(\Omega_m,\Omega_{\Lambda},\Omega_b,h,\sigma_8,n_s) =
(0.279,0.721,0.046,0.701,0.817,0.96)$. The simulation was performed
in a box of 2000 $h^{-1}{\rm Mpc}$ on a side, 
containing $512^3$ particles whose mass is $4.6\times 10^{12}\ 
h^{-1} M_\odot$. 
The measurements were made from the simulation output at
$z=0.5$.
To account for the discrete nature of $N$-body measurements,
 separation $r$ in the following refers to two particles having a
 separation between
$(r-2,r+2)$ in the unit of $h^{-1}{\rm Mpc}$.
For the theoretical model we apply the same separation selection in the
dirac-delta function of $r$. 
The comparisons of the velocity PDF of $v\lpa$ and $v\lpe$ are
presented in the next two subsections.
We will then discuss the implications of the
comparisons in the next section.

\subsubsection{Parallel to the line of separation: $p(v\lpa ; r, f_{nl})$}
\begin{figure}
\centering
  \subfloat[r=4 ${\rm Mpc}\ h^{-1}$]{
      \label{fig:pdfvll_r4}
      \includegraphics[width=0.44\linewidth]{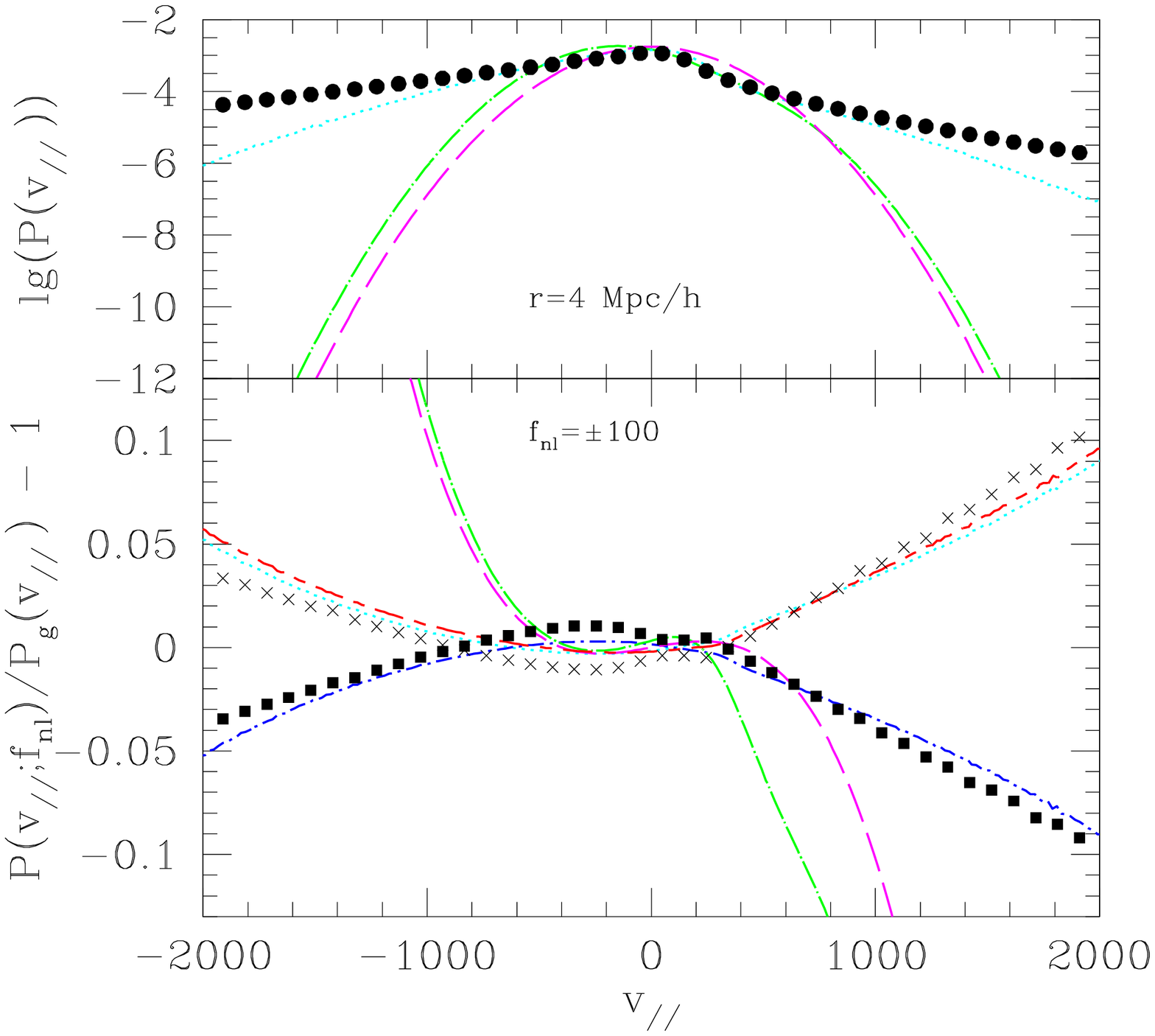}}
   \hspace{0.07\linewidth}
  \subfloat[r=8 ${\rm Mpc}\ h^{-1}$]{
      \label{fig:pdfvll_r8}
      \includegraphics[width=0.44\linewidth]{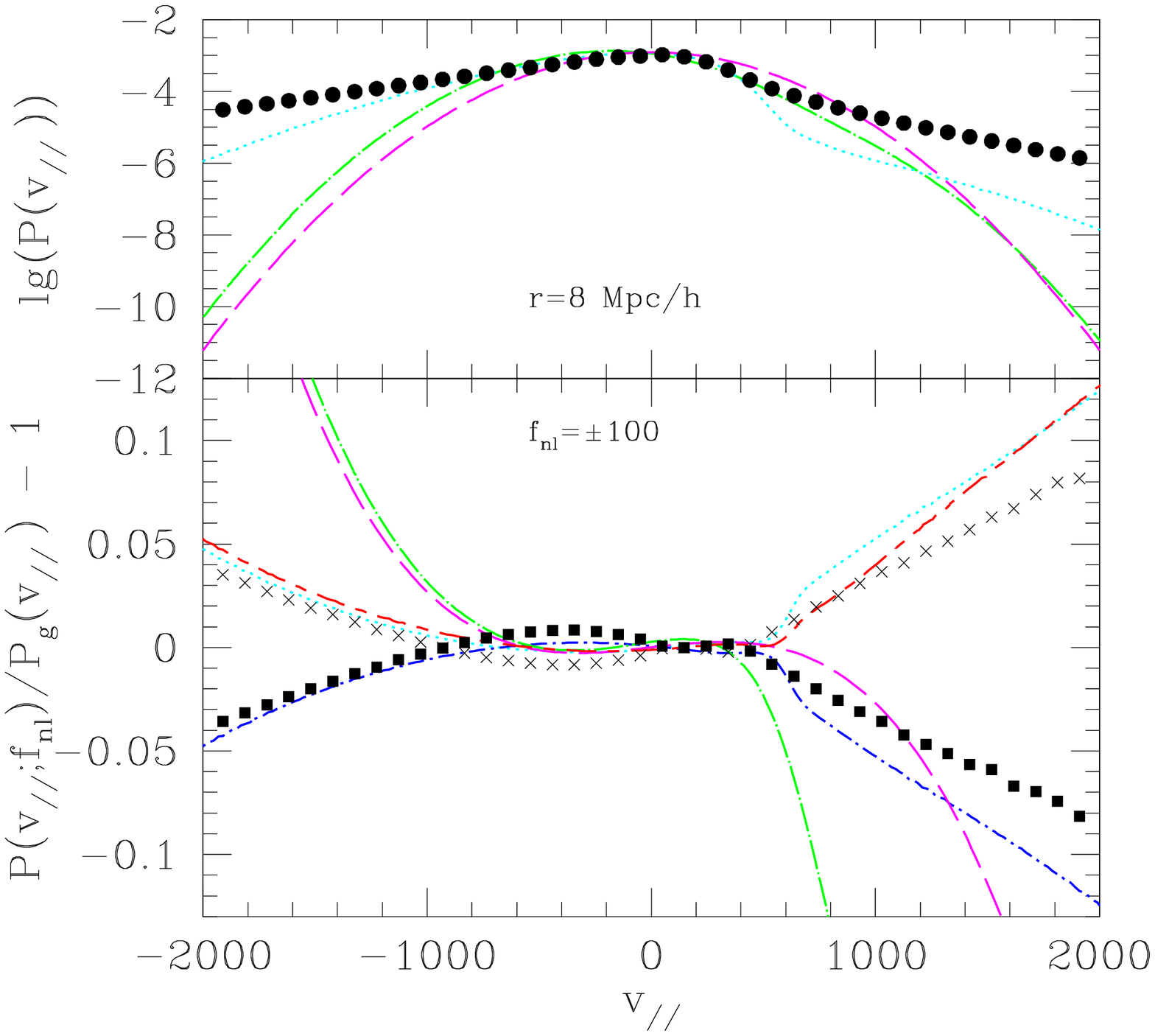}} \\[10pt]
  \subfloat[r=12 ${\rm Mpc}\ h^{-1}$.]{
      \label{fig:pdfvll_r12}
      \includegraphics[width=0.44\linewidth]{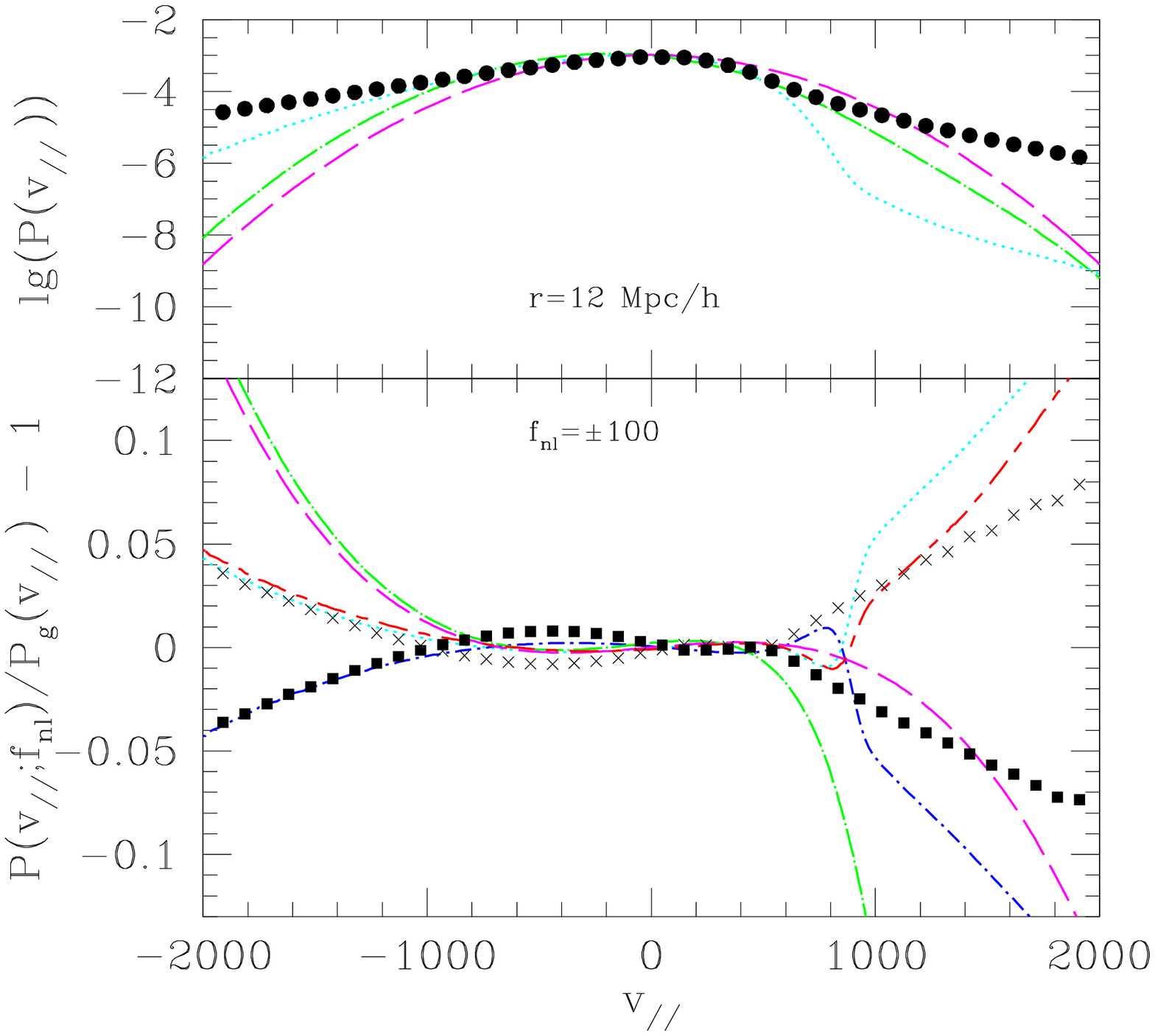}}
   \hspace{0.07\linewidth}
  \subfloat[r=50 ${\rm Mpc}\ h^{-1}$]{
      \label{fig:pdfvll_r50}
      \includegraphics[width=0.44\linewidth]{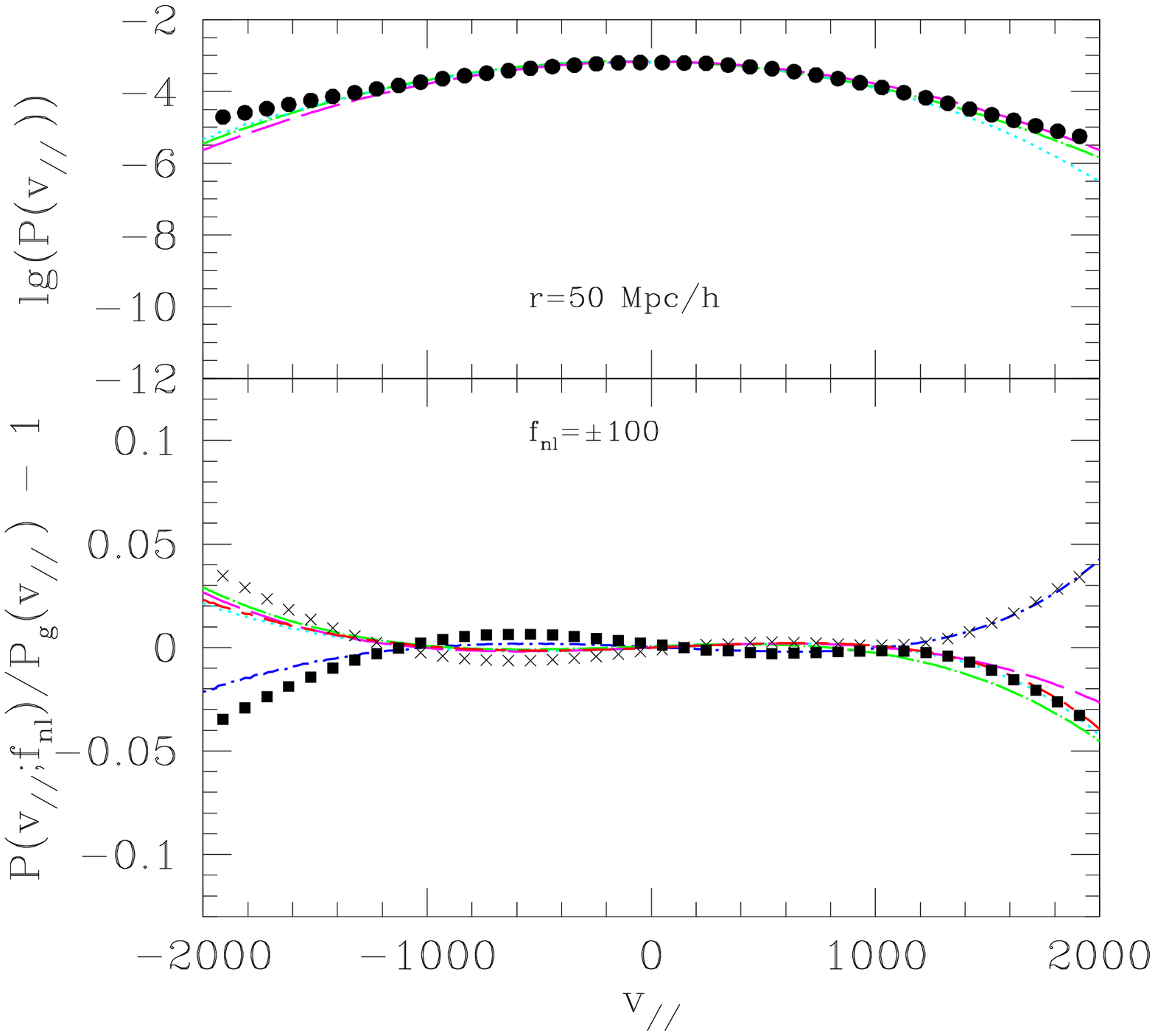}}
 \caption{Pairwise velocity (parallel to line of separation) PDF at
   different separations. 
   Negative velocity corresponds to particles infalling while positive
   velocity indicates particles moving away from each other.
   Upper panels show the pairwise velocity PDF when $f_{nl} = 0$:
   solid symbols are measurements from the $N$-body simulation, 
   cyan dotted curves are the velocity difference PDF from the evolution model
   based on the Zeldovich
   approximation (equation~\eqref{eqn:evolvedpdf}), 
   green dot-long dashed curves are the linear pairwise
   velocity ($q_0$ in equation~\eqref{eqn:q0p0}), and 
   magenta long dashed curves are the linear
   velocity difference PDF (equation~\eqref{eqn:pdfgvw}).
   Lower panels show the ratios of the PDF for $f_{nl} \pm 100$ to
   the associated PDF for $f_{nl} = 0$: crosses ($f_{nl}=100$) and
   solid squares ($f_{nl}=-100$), cyan dotted curves ($f_{nl}= 100$)
   and blue dot-short dashed curves ($f_{nl}=-100$) are predictions of
   equation~\eqref{eqn:evolvedpdf},
   green dot-long dashed curves are the ratios for the 
   linear pairwise PDF ($f_{nl}=100$), and 
   magenta long dashed curves are the ratios for the 
   linear velocity difference PDF ($f_{nl}= 100$). 
    Red short-long-dashed curves show
  the prediction from equation~\eqref{eqn:evolvedpdf}, neglecting
  third order moments involving velocity perpendicular to line of
  separation (see the lower right panel of figure~\ref{fig:pdfvlinearVW}).
    }
 \label{fig:pdfvll}
\end{figure}

Figure~\ref{fig:pdfvll} shows the PDF comparisons of the various analytical
predictions to the $N$-body measurements for the relative velocity parallel
to the line of separation for 4 different separations (4, 8, 12, 50 
$h^{-1}{\rm Mpc}$).
The upper panel in each subfigure shows the PDF profile for $f_{nl}=0$: 
solid symbols are the
$N$-body measurements, cyan dotted curves are the analytical predictions of
equation~\eqref{eqn:evolvedpdf} (marginalized over $v\lpe$), green
dot-long dashed (mass weighted) and magenta long dashed (uniform
weighted) curves are the linear theory predictions respectively.

The predictions from our analytical evolution model give good matches
to the $N$-body measurements at all scales we investigated, except at
large outgoing velocities (i.e. $v\lpa \gg 0$). 
The agreements with the $N$-body measurements are better in small
separations (4 and 8 $h^{-1}{\rm Mpc}$) than large separation (12
$h^{-1}{\rm Mpc}$). 
The analytical predictions of our model are able to describe the 
 high infalling velocity regimes
($v\lpa < -1000$). 
This is very encouraging since one may expect that high
velocity regions are not described by our simple model. 
The analytical model also predicts 
knee-shape transitions at around $300$, $600$,
and $900 {\rm km/s}$ for $r= 4$, $8$, $12\ h^{-1}{\rm Mpc}$
respectively.
Similar but less significant 
knee-shape changes can be seen from the $N$-body measurements 
around the corresponding velocities.
In the high outgoing velocity regions the match is not as good as the
infalling regimes. We believe the disagreement 
in the outgoing velocity regime as compared to the infalling regime is
due to non-linear evolution of velocity that is not described by our
current model -- originally infalling pairs eventually become
outgoing (see equation~\eqref{eqn:vll}) and non-linear interactions
between particles in close contact are not described by the 
Zeldovich Approximation.

On the other hand, the linear theory predictions do not fare as well
as the analytical model's.
At small separations the predictions of the  
linear velocity difference PDF do not match the $N$-body measurements --
the magneta curves miss both the extrema and the peaks of the PDF
at separation $r=$ 4, 8, and 12 $h^{-1}{\rm Mpc}$. 
This is consistent with previous studies that the linear theory 
uniform weighted velocity PDF 
does not agree with the velocity correlation in the
parallel to the line of separation direction.
On the contrary the mass weighted linear predictions provide 
reasonable matches to the $N$-body measurements near the peaks of the
PDF at different separations.
It confirms the finding of \citet{sz09}: 
the velocity correlation can be described by the linear theory when the
mass weighting is taken into account. 
The matching of the peaks of the PDF
guarantees the velocity
correlations, which is equivalent to the expected value of the
pairwise velocity PDF, would roughly agree. 
However the mass weighted linear
predictions does not match the $N$-body measurements when $|v\lpa|$ is
of the order of a few hundreds ${\rm km/s}$. 
The disagreement is getting worse as the separation decreases.
At large separation (50 $h^{-1}{\rm Mpc}$), 
the two linear theory predictions
are very similar and they describe the $N$-body measurements well.
They are very closed to the prediction of the analytical model 
and the match to the measurements is only slightly worse at high
velocity regimes.

The lower panels of figure~\ref{fig:pdfvll} show the ratios
(substracted by unity) of the
pairwise velocity PDFs for $f_{nl} = \pm 100$ to the associated
PDFs for $f_{nl}=0$ at different separations.
 Measurements from the $N$-body simulations are
represented by crosses ($f_{nl}=100$) and squares ($f_{nl}=-100$). 
Other curves in the lower panels show the PDF ratios of 
different analytical predictions: 
cyan dotted ($f_{nl}=100$) and blue dot-short dashed ($f_{nl}=-100$)
curves are the predictions of the evolution model based on the 
Zeldovich Approximation 
(equation~\eqref{eqn:evolvedpdf}); red short-long-dashed curves show
the similar predictions 
for $f_{nl}=100$ 
but setting $\lrangle{v\lpa \mathfrak{v}\lpe^2}=0$
(see the lower right panel of figure~\ref{fig:pdfvlinearVW} 
for the corresponding linear theory comparison at
$r=8\ h^{-1}{\rm Mpc}$). 
The other two curves are the linear theory predicted ratios 
for $f_{nl}=100$: the magneta long dashed curves are the
uniform weighted 
velocity predictions and 
the green dot-long dashed curves are
the mass weighted 
velocity predicted ratios.

For the range of separations we investigated the change in the pairwise
velocity PDF due to 
$f_{nl}=\pm 100$ is at most 5\% in the infalling velocity regime and 
10\% in the outgoing velocity regime. 
In contrary to the linear theory
predictions (see figures~\ref{fig:pdfvlinearVW} and
\ref{fig:pdfvlinearMW}, and the magenta long dashed as well
as the green dot-long dashed curves), the effect of positive $f_{nl}$ 
enhances the probability of having pairs in \emph{both} the infalling
and the outgoing high velocity ends. 
This  is true 
for all the separations we look at, from 4
to 50 $h^{-1}{\rm Mpc}$ -- 
scales in which the linear theory is believed to be valid.
We also check using a simulation with a smaller box ($1\
h^{-1}{\rm Gpc}$ on a side) to 
estimate the box size effect: shrinking
the box volume to one-eighth of its size changes the PDF ratios by at most 2\% at
$v\lpa = 2000\ {\rm km/s}$.

The predictions of our evolution model, regardless of whether
one explicitly set $\lrangle{v\lpa \mathfrak{v}\lpe^2}=0$
, match the $N$-body measurements reasonably well 
when $r=4\ h^{-1} {\rm Mpc}$. 
The predicted ratios from the analytic model match the measurements
within 2\% across the velocity range we look at.
It is remarkable to have such agreement at velocity as high as 
$\sim 2000{\rm km/s}$.
The predicted ratios also match the transition of the PDF from the
infalling to the outgoing regime:  
for $f_{nl}=100$ 
it gradually decreases from positive
to negative, then changes direction and become positive again. 
The same transition is also observed at larger separations and the
model's predictions are able to match the $N$-body measurements.
In contrast the ratios of the linear theory predictions (both the
uniform and the mass
weighted) fail to match 
the $N$-body measurements
in both infalling and outgoing velocity regimes. 
While the linear theory predicted ratios still have the same trend
as the $N$-body measurements in the infalling velocity regime 
(but the differences between the predictions and the measurements are large), 
the predicted ratios from linear theory 
are opposite to the
measurements in the outgoing velocity regime (the same is also
observed at larger separations).
Hence the linear theory cannot be used to predict the effect of
$f_{nl}$ on the pairwise velocity PDF.

At larger separations 
the agreements in the ratios of the PDF 
between our analytical model predictions
and the $N$-body measurements are less impressive:
while 
the predicted ratios 
still agree well with the
measurements in the infalling velocity regime,
the differences in the outgoing velocity regime are getting bigger as the
separation increases.
Nonetheless the predictions still qualitatively match the measurements
at $r= 8 $ and $12 \ h^{-1}{\rm Mpc}$. 
At $r= 50\ h^{-1}{\rm Mpc}$, on the other hand,
the analytical predictions converge to the linear theory
predictions and show opposite trends to the measurements. 
In the outgoing velocity region,
there are some interesting transitions in the analytical model
predictions: 
at $r = 12\ h^{-1}{\rm Mpc}$ and $f_{nl} = 100$,
the predicted ratios (the cyan as well as red curves) first show a decrement 
at small $v\lpa$. They then turn around near 
$v\lpa = 800 {\rm km/s}$ and keep increasing for larger
$v\lpa$. 
Such transitions are not seen in the $N$-body measurements.

We will leave the discussion on the implications for the discrepancies between the linear
theory predictions and the $N$-body measurements at all separations 
as well as the mismatch in the PDF ratios between our analytical
models and the measurements at large separations in the next section.

\subsubsection{Perpendicular to the line of separation: $p(v\lpe ;
  r,f_{nl})$}
\begin{figure}
\centering
  \subfloat[r=4 ${\rm Mpc}\ h^{-1}$]{
      \label{fig:pdfvl__r4}
      \includegraphics[width=0.45\linewidth]{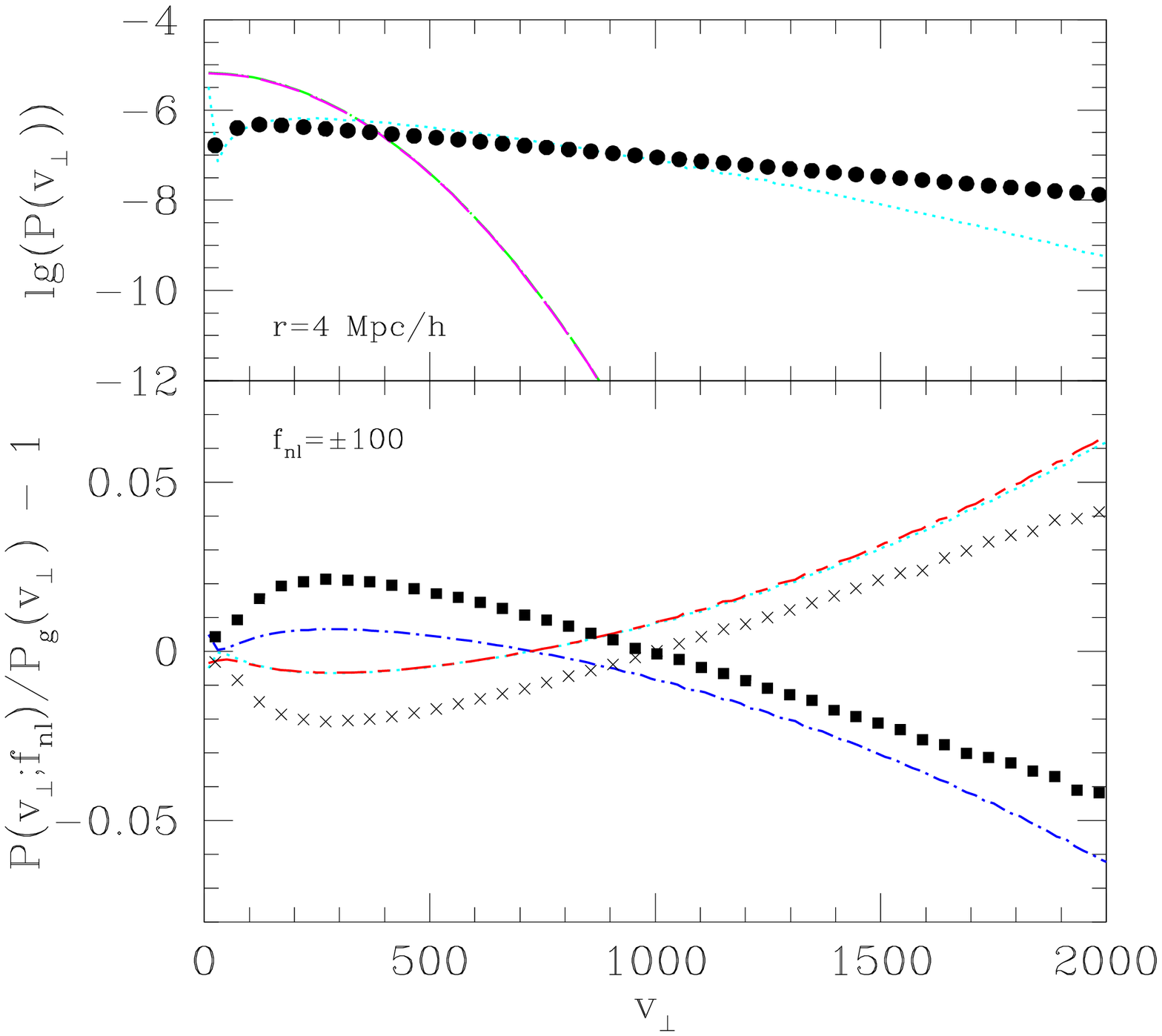}}
   \hspace{0.07\linewidth}
  \subfloat[r=8 ${\rm Mpc}\ h^{-1}$]{
      \label{fig:pdfvl__r8}
      \includegraphics[width=0.45\linewidth]{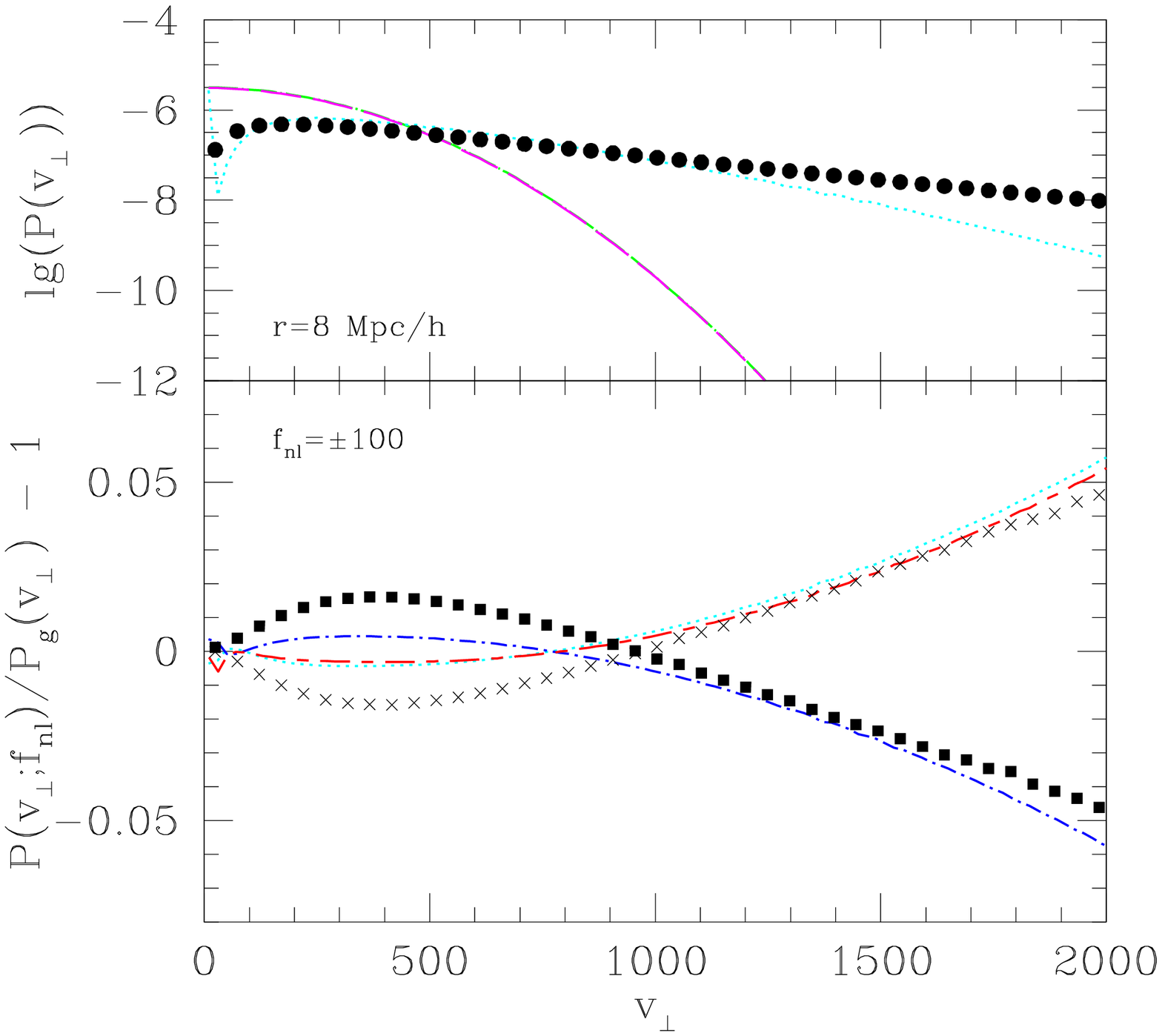}} \\[10pt]
  \subfloat[r=12 ${\rm Mpc}\ h^{-1}$]{
      \label{fig:pdfvl__r12}
      \includegraphics[width=0.45\linewidth]{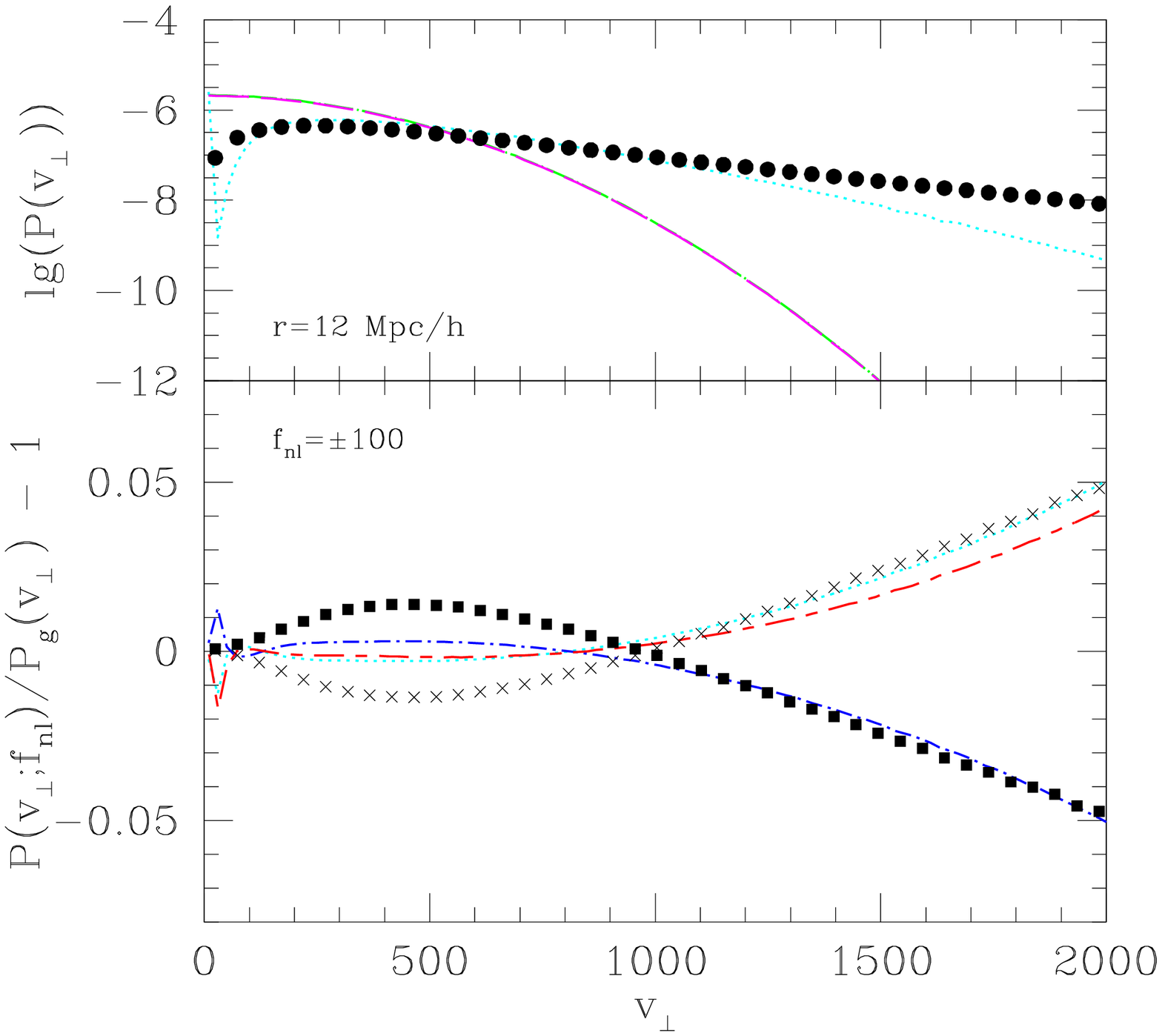}}
   \hspace{0.07\linewidth}
  \subfloat[r=50 ${\rm Mpc}\ h^{-1}$]{
      \label{fig:pdfvl__r50}
      \includegraphics[width=0.45\linewidth]{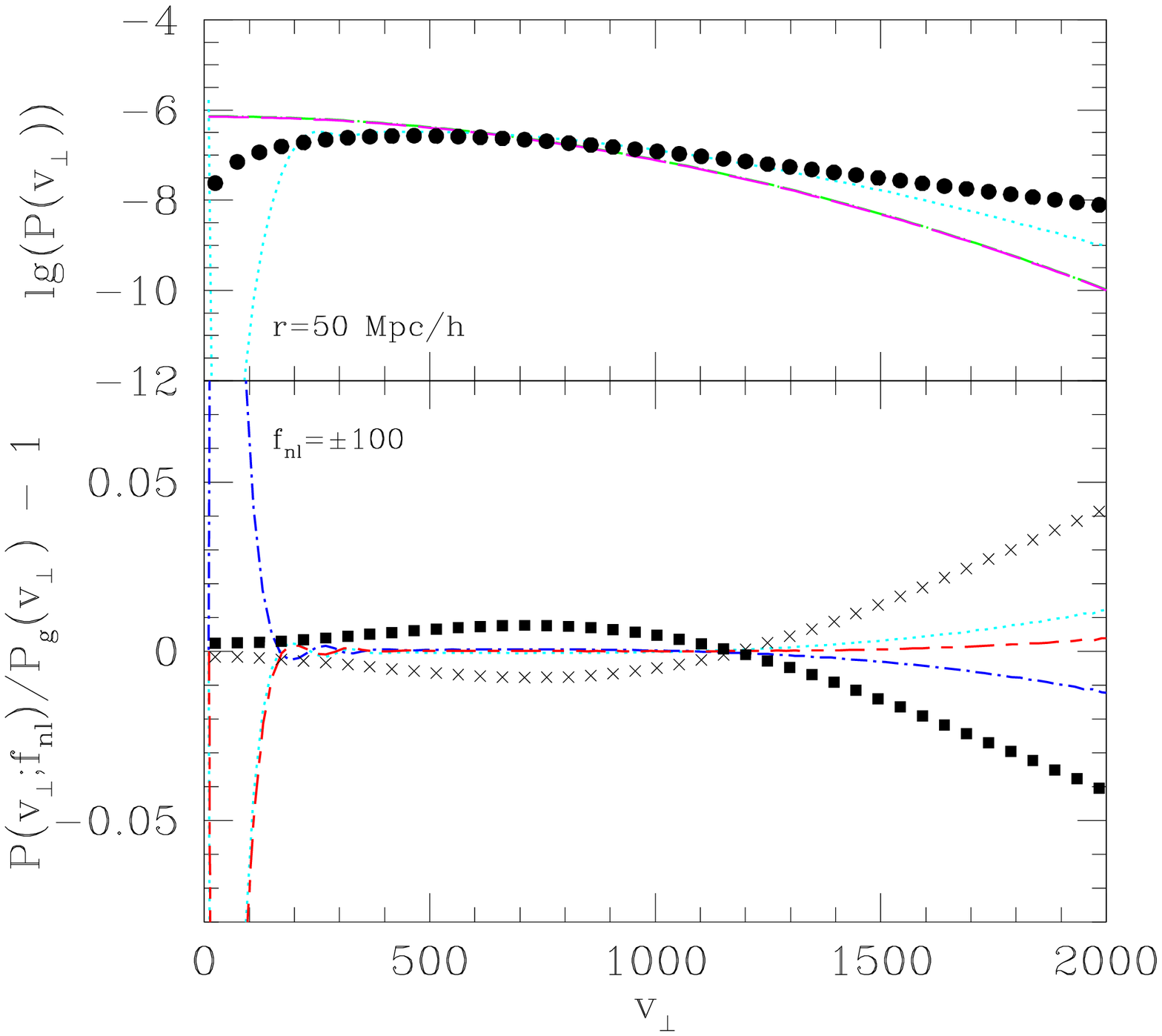}}
 \caption{Pairwise velocity (perpendicular to line of separation) PDF at
   different separations. 
   Upper panels show the PDF profile when $f_{nl} = 0$:
   solid symbols are measurements from the $N$-body simulation, 
   cyan dotted curves are the prediction from the evolution model
   based on the Zeldovich
   Approximation (equation~\eqref{eqn:evolvedpdf}), 
   green dot-long dashed curves are the linear pairwise
   velocity ($q_0$ in equation~\eqref{eqn:q0p0}), and 
   magenta long dashed curves are the  linear
   velocity difference PDF (equation~\eqref{eqn:pdfgvw}).
   Lower panels show the ratios of the PDF when $f_{nl} \pm 100$ to
   the associated PDF for $f_{nl} = 0$: crosses ($f_{nl}=100$) and
   solid squares ($f_{nl}=-100$) are measurements from $N$-body simulation, 
    cyan dotted curves ($f_{nl}= 100$)
   and blue dot-short dashed curves ($f_{nl}=-100$) are predictions of
   equation~\eqref{eqn:evolvedpdf}.
     Red short-long-dashed curves show
  the prediction from equation~\eqref{eqn:evolvedpdf}, neglecting
  third order moments involving velocity perpendicular to line of
  separation (see the lower right panel of figure~\ref{fig:pdfvlinearVW})
    }
 \label{fig:pdfvl_}
\end{figure}

Figure~\ref{fig:pdfvl_} shows the comparisons of the PDFs of 
the pairwise velocity perpendicular to the line of separation for
different separations. The upper panels show the PDF at $z=0.5$ 
for $f_{nl}=0$ and the lower panels show the ratios of the PDF 
for $f_{nl}=\pm 100$ to the associated PDF for $f_{nl}=0$. 
The symbol and curve labelings are the same as in
figure~\ref{fig:pdfvll}, but we do not show the PDF ratio predictions of 
the linear theory
in the lower panels since the linear theory predicts no effect of
$f_{nl}$ in the marginalized PDF, that is 
$p(v\lpe ;r,f_{nl})/p(v\lpe ;r,f_{nl}=0) =1$. 

The predictions of the PDF profile from our analytical model
(equation~\eqref{eqn:evolvedpdf}) match the $N$-body measurements reasonably
well, except around $v\lpe = 0$ at $r=50 \ h^{-1}{\rm Mpc}$. 
The predictions agree with the measurements
in a wide range of velocity ($200-1500 {\rm km/s}$), as well as the
dips near $v\lpe =0$. 
At higher $v\lpe$ the model predictions always underestimate the PDF.
The linear theory predictions, on the other hand, fail to match the
measured PDF for $r=4$, $8$, and $12\ h^{-1}{\rm Mpc}$. 
At $r=50\ h^{-1}{\rm Mpc}$ the linear theory prediction agrees with
the $N$-body measurements for $v\lpe = 300-1200 {\rm km/s}$, but it
fails to predict the dip near $v\lpe = 0$.
Note that there is no difference between the mass weighted and the uniform
weighted linear theory predictions since the correction term due to
mass weighting does not depend  on $v\lpe$
(equation~\eqref{eqn:q0p0}).

The effect of primordial non-Gaussianity in the pairwise velocity PDF 
in the perpendicular to the line
of separation direction is not as strong as the signature in the parallel to
the line of separation direction: the change in the PDF is top at  7\% at
$v\lpe = 2000 {\rm km/s}$. The ratio first shows a decrement
(increment) 
at small $v\lpe$ for $f_{nl}=100 \ (-100)$. 
It then gradually
switches direction and at high $v\lpe$ show an increment (decrement).
This gradual switch from decrement to increment 
is most significant at small separation.
Our analytical model is able to predict the
change in the PDF due to primordial non-Gaussianity qualitatively. 
Its prediction matches the gradual change of 
the PDF ratios (from decrement to enhancement 
for $f_{nl}=100$, opposite for $f_{nl}=-100$) 
and provides a rough approximation on 
the velocity where the crossing across unity  occurs.
However the predictions 
always underestimate 
the PDF ratios in low $v\lpe$ regime. 
At higher $v\lpe$ the model prediction on the PDF ratio
matches the $N$-body measurements
very well at $r=8$ as well as  $12 \ h^{-1}{\rm Mpc}$;
the predicted ratio overestimates the change at $r = 4\ h^{-1}{\rm
  Mpc}$ but underestimates at  $r = 50\ h^{-1}{\rm Mpc}$.
The analytical model predicts the effect of
$f_{nl}$ becomes weaker at some fixed $v\lpe$ for larger $r$, 
the measurements from the $N$-body simulations, on the contrary, 
remain roughly the same 
at various separations.
The measured PDF ratios at high $v\lpe$
only depends on the scale of separation very weakly.
In addition there are spikes near $v\lpe = 0$ in the theoretical
predicted ratios for $r = 12$ and $50\ h^{-1}{\rm Mpc}$,
 but these are not seen in the $N$-body simulations.

The linear theory predictions fail to match the $N$-body measurements
 at all separations -- it predicts no change in the pairwise
velocity PDF in the perpendicular to the line of separation direction. 
The $N$-body measurements, on the other hand, has several percent-level
change in the PDF throughout the velocity range we look at.

\section{Discussions}
\label{section:discussion}
We have studied how primoridal non-Gaussianity  affects 
the probability density function of the pairwise
velocity. Our study complements earlier studies on the signatures of
primordial non-Gaussianity on other large-scale structure probes 
which utilize the modification in the density field (for example the PDF
of density field, halo/void abundances, scale dependent halo bias).
We have shown that primordial non-Gaussianity models with a non-vanishing 
bispectrum result in two non-vanishing third-order  connected 
moments of the peculiar velocity ($\lrangle{v\lpa^3}$ and 
$\lrangle{v\lpa \mathfrak{v}\lpe^2}$).
We adopt the local $f_{nl}$ type primordial non-Gaussianity 
as an example to compute the modification in the linear velocity
difference PDF. 
Following  suggestions by earlier studies,
we investigate the effect of $f_{nl}$ on the 
linear pairwise velocity PDF. 
Primordial non-Gaussianity induces correlations between velocities in
the parallel and the perpendicular to the line of separation
directions. Hence both the uniform weighted and the mass weighted linear 
velocity PDF can no longer be written as a product of the corresponding
univariate PDFs. 
Futhermore the change in the linear velocity PDF 
 is degenerated in the $(v\lpa, v\lpe)$ plane.

Next we have developed an analytical model to describe the evolution of the
 velocity PDF. It employs  the Zeldovich Approximation
which assumes the comoving velocity remains unchanged.
We then compared the pairwise velocity PDF measured from a 
$N$-body simulation with the predictions from our analytical model
at various  separations, ranging from 
$4$ to $50\ h^{-1}{\rm Mpc}$.
Our model predictions agree well with the PDF profiles 
measured from the $N$-body simulation at various separations when $f_{nl}=0$ 
in both the parallel and in the perpendicular to the
line of separation directions.

Linear theory fails in matching the $N$-body measurements:
both the linear velocity difference PDF as well as the linear pairwise
velocity PDF fail to predict the PDF profiles and the change in 
the PDF due to a non-zero $f_{nl}$.
The linear theory prediction does not take into account the movement
of particles: pairs separated by $r$ having a relative velocity 
$(v\lpa, v\lpe[a],v\lpe[b])$ must have had different separations at
earlier redshift. The relative velocities at earlier redshifts
 also change in general due to the difference in the particles' positions.
In other words linear theoy is a good approximation 
only when the displacement of
the particles is small compared to the separation between them.
For example the displacement associated with a relative velocity
$v=1000\ {\rm km/s}$ is approximately $10\ h^{-1}{\rm Mpc}$ at $z=0.5$
if the velocity remains unchanged. 
Therefore, only for large enough separations ($r \gg 10 h^{-1}{\rm Mpc}$) 
the linear theory can approximate the PDF.
The initial separation and the initial relative velocity determine
the evolved separation and the pairwise velocity -- 
equations~\eqref{eqn:rsq} to \eqref{eqn:vl_} give an example 
of such an evolution model using the Zeldovich approximation. 
In general the mapping from the initial parameters 
$\{r^i,v\lpa^i,v\lpe^i \}$ to the evolved parameters 
$\{r,v\lpa,v\lpe \}$ is non-linear and more than one set of
$\{r^i,v\lpa^i,v\lpe^i \}$ would evolve to  $\{r,v\lpa,v\lpe \}$.
This analytical model provides results that match the measurements
much better than the linear theory.

Including evolution 
is also important in
explaining the change in the PDF due to $f_{nl}$.
The ratios of the pairwise velocity PDF in the parallel to the line of
separation direction can be used to reveal the distribution of pairs
with different $\{r^i,v\lpa^i,v\lpe^i \}$ having evolved to the same
$\{r,v\lpa,v\lpe \}$.
Recall that the linear theory predicts enhancement in the probability
occurs in high velocity infalling pairs and low velocity outgoing
pairs when $f_{nl}>0$. 
On the other hand, the $N$-body measurements always have 
increment in the probability for outgoing pairs.
The analytical model provides an algorithm to explain this
discrepancy:
pairs initially infalling becomes outgoing when they
cross each other on the projection on the line of separation.
As a result outgoing pairs would be either originally infalling or
outgoing pairs. The converse is not true 
because infalling pairs can
only come from pairs that approach each other at earlier redshift.
The originally low velocity outgoing pairs, having a higher
probability for $f_{nl}=100$, by itself cannot explain the observation from
$N$-body measurements  as the
change in the velocity difference PDF associated with 
these pairs is relative small
(recall that the change in the PDF goes as $v\lpa ^3 - 3v\lpa$ 
and $v\lpe^2 -1$) and cannot account for the level of change in the PDF
measured in the simulations.
As a result the PDF ratio in the outgoing velocity regime has to be
dominated by original infalling pairs.

Our analytical model provides an approximation for the evolution of
the velocity difference PDF and it successfully predicts the
modification in the velocity PDF due to $f_{nl}$ qualitatively. 
The Zeldovich Approximation, on which the analytical model based, 
does not take into account several important features -- 
it assumes the comoving velocity remains unchanged. This is a good
approximation when the separation of the 
pair in consideration
is big. This assumption may not hold 
when the pair 
crosses on
the projection along the line of separation. 
Non-linear model is
required to describe the evolution of the relative velocity in this
cases. As a result the velocity PDF in the regime where cross-over
dominates may not be well described by our analytical model.

Another area  our analytical model does not describe is the virial
motion of particles residing inside dark matter halos. 
While it is not expected to have pairs 
residing within the 
same dark matter halo for the separation we considered in this study, 
virial motions would certainly affect the distribution of the pairwise
velocity.

Notice that our analytical model only describes the evolution of the
velocity difference PDF -- it is the evolved uniform weighted velocity
PDF. In practice it is not what we measured from
the $N$-body simulation -- it is a (evolved) pair weighted velocity
PDF. 
The comparisons with $N$-body measurements show
that the evolved velocity difference PDF matches the measurements
better than the linear  pairwise velocity PDF. 
Modeling of the non-linear pairwise velocity PDF requires the
correlation between the non-linear density contrast and the evolved
velocity field. These are not described in the our model and will be
left for future study.

It is quite unexpected that the agreement of the analytical model 
with the $N$-body measurement is better at small separation than at
large separation. One speculation for that is the simulation
measurement is affected by the finite box size effect for big $r$ -- we did 
check the effect of finite box size by measuring the pairwise velocity
PDF from a simulation ran in a box with 1 $h^{-1}{\rm Gpc}$ on a
side. The resulting PDF at $r = 50 \ h^{-1}{\rm Mpc}$ agrees
qualitatively with the corresponding PDF in the main text.
Hence we believe the disagreement in the PDF ratio at $r = 50 \
h^{-1}{\rm Mpc}$
is not due to 
the effect of finite box size.
Another possibility is the discrepancy is due to the non-linear
interactions -- here we give two approaches to explain qualitatively  how
the non-linear interactions affect the velocity PDF. Both are
related to the formation of non-linear structures and is not described
by the Zeldovich Approximation.

The Zeldovich Approximation assumes the comoving velocity is constant
throughout the evolution and it is not a good approximation when
non-linear structures (pancake, filament, and halos) form. 
If the Zeldovich Approximation were the correct model
that completely described the evolution, the positive velocity in
the evolved velocity difference PDF would be dominated by originally
infalling pairs only 
when $v\lpa \gtrsim v_0$, where $v_0$ corresponds to the velocity that has
a displacement approximately equal to the separation $r$.
This scale also associates with the knee-shape transition in the PDF
profiles in the upper panels of figure~\ref{fig:pdfvll}.
For $v\lpa < v_0$ the Zeldovich Approximation 
predicts originally outgoing pairs should dominate the PDF.
Measurements from $N$-body simulation, on the other hand, 
show no such transition. The knee-shape transition in the PDF profiles
(upper panels) is also much weaker compared to the analytical
predictions.
Non-linear evolution models, such as the adhesion model, may be able
to explain the difference between the measured and the predicted PDF
-- while the Zeldovich approximation predicts the velocity PDF is
dominated by originally infalling pairs only when $v\lpa > v_0$, 
the frictional dragging term in the adhesion model 
would damp the
velocity, making $v_0 < v'\lpa \rightarrow v\lpa < v_0$. 
As a result the velocity PDF would be dominated by originally
infalling pairs at a smaller $v\lpa$ compared to the Zeldovich
Approximation prediction. Notice that this effect will be most
significant for big $r$ as the dragging term is proportional to $v$.

The halo model provides another explanation for this discrepancy: in
the halo model all particles are residing in halos, each of them 
acquires its peculiar velocity from two
contributions -- the virial motion inside the parent halo as well as
the bulk movement of the parent halo
\citep{haloreview,sdvel01,shds01}. 
Hence the relative velocity between
pairs residing in different halos (this corresponds to the separations we
studied in the current study)  can be separated into two parts: the randomly-oriented
velocity difference due to the corresponding virial motion and the
relative velocity of the two parent halos. 
The dispersion of the virial motion 
depends on the mass of the parent
halo: more massive halos have a larger dispersion;
the bulk velocity of the parent halo depends only weakly  on mass.
The bulk velocity can be approximated by the linear theory \citep{haloreview}.
At relatively small separations, the two contributions are comparable; 
at big separations, on the other hand, the two-point correlation
function is weak and the velocity is dominated by the virial motion term.
It is a well-known fact that there are more massive halos for models
with $f_{nl}>0$ \citep[see, for example, ][]{lsdfnl}, hence the
velocity dispersion is bigger when $f_{nl}>0$. This effect is most
significant in the pairwise velocity PDF for big $r$ since at such
separation the dominant velocity difference is the random-oriented
virial velocity difference. The net result is having a higher
probability when $|v\lpa|$ is big but lower probability when $|v\lpa|$
is small.
Our current analytic model does not include the virial motion and as a
result it is not able to describe the change in the velocity PDF for
large $r$.
Note that this picture naturally predicts the discrepancy should 
first show in the outgoing velocity regime -- clustered halos
attract each other and hence the relative bulk velocity is
always infalling when projected on the line of separation of the pairs
in consideration.
\begin{figure}
\centering
\includegraphics[width=0.7\textwidth]{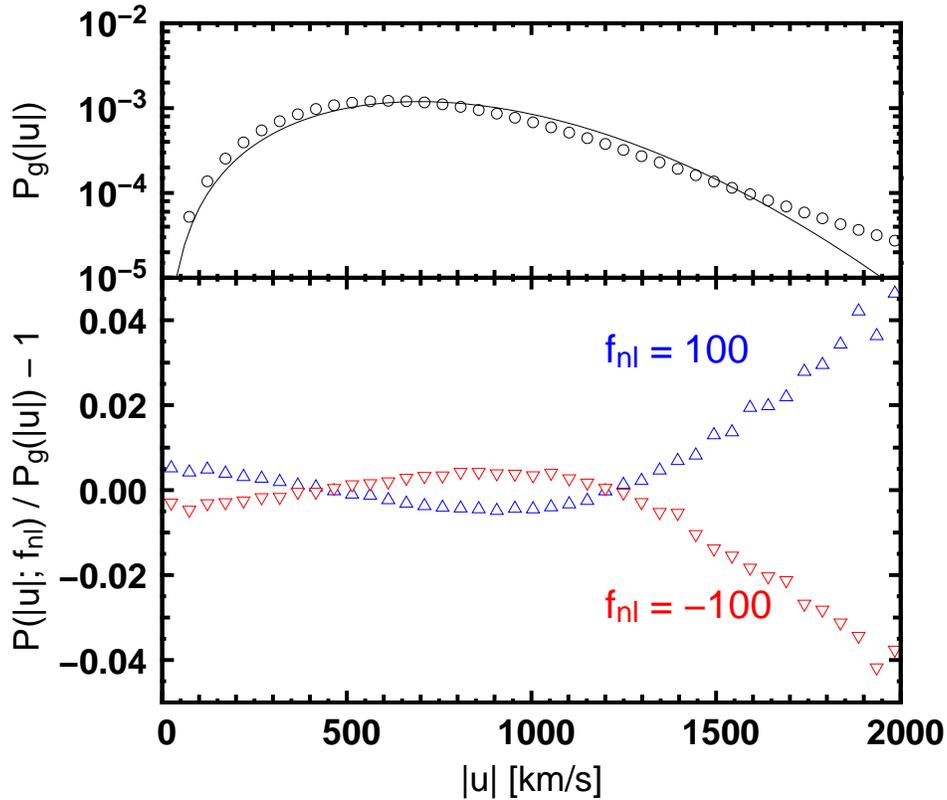}
\caption{The 1-point PDF of peculiar velocity: the upper panel shows
  the PDF profile of the peculiar velocity when $f_{nl}=0$. 
  The open circles show the measurement from $N$-body simulation and
  the solid curve shows the Maxwellian distribution with $\sigma$
  equal to the velocity dispersion measured from the simulation. The
  lower panel shows the ratio of the PDF for $f_{nl} = \pm 100$ to
  $f_{nl}=0$: upward pointing triangles are the ratio for $f_{nl}=
  100$; downward pointing for $f_{nl}= -100$. 
Particles with large velocities ($> 1000 {\rm km/s}$)
are mostly populated in massive halos.}
\label{fig:oneptPDF}
\end{figure}
Figure~\ref{fig:oneptPDF} shows the change in the PDF of peculiar
velocity for models with $f_{nl}\neq 0$. The increase in velocity
dispersion  and the change in PDF at high $v$ when $f_{nl} =100$ 
supports this halo model picture in
explaining the discrepancy at large separation due to the increase in
the magnitude of the virial motion in $f_{nl} (>0)$ models.

It is out of the scope of the current study to include these
non-linear effect -- this will be left for future study. 
Our current analytic model is able to predict the modification in the
pairwise velocity PDF within a few percent level at quasi-linear
regime (from $4-12\ h^{-1}{\rm Mpc}$).

While this study focuses on the pairwise velocity of PDF of dark matter,
we are in the process to extend the current work to study the effect
of primordial non-Gaussianity on the pairwise velocity of biased
tracers.
We are also extending our study to examine the effect of primordial
non-Gaussianity on the redshift space distortion. 
The $N$-body measurements of the pairwise velocity PDF 
show clearly both the parallel and the perpendicular to the line of
separation directions have primordial non-Gaussianity imprints.
The effect can be seen in separation as large as 
$r=50\ h^{-1}{\rm Mpc}$. 
Our analytical model
improves earlier work \citep{lsdfnlred,schmidt10} --
it includes
the change in the pairwise velocity PDF in both the parallel and the
perpendicular directions in the analysis as well as the evolution of
the pairwise velocity PDF. 
As a result it provides a much better match to the $N$-body measurements
compared to the linear theory. 
This enables our analytical model to improve the estimation 
of the signature of primordial non-Gaussianity in the redshift space distortion.


\section*{Acknowledgements}
TYL would like to thank Ravi Sheth, Robert Smith, and Vincent
Desjacques for the discussion on peculiar velocity field.
This work was supported by World Premier International Research Center
Initiative (WPI Initiative), MEXT, Japan. TYL is also supported by a
kakenhi grant (Grant-in-Aid for Young Scientists (B) -- 22740149).
T. N. is supported by a Grant-in-Aid for Japan Society for the 
Promotion of Science (JSPS) Fellows.


\begin{thebibliography}{}

\bibitem[\protect\citeauthoryear{{Afshordi} \& {Tolley}}{{Afshordi} \&
  {Tolley}}{2008}]{at08}
{Afshordi} N.,  {Tolley} A.~J.,  2008, \prd, 78, 123507

\bibitem[\protect\citeauthoryear{{Bernardeau}, {Colombi}, {Gazta{\~n}aga} \&
  {Scoccimarro}}{{Bernardeau} et~al.}{2002}]{ptreview}
{Bernardeau} F.,  {Colombi} S.,  {Gazta{\~n}aga} E.,    {Scoccimarro} R.,
  2002, \physrep, 367, 1

\bibitem[\protect\citeauthoryear{{Bhattacharya} \& {Kosowsky}}{{Bhattacharya}
  \& {Kosowsky}}{2007}]{bk07}
{Bhattacharya} S.,  {Kosowsky} A.,  2007, \apjl, 659, L83

\bibitem[\protect\citeauthoryear{{Bhattacharya} \& {Kosowsky}}{{Bhattacharya}
  \& {Kosowsky}}{2008}]{bk08sz}
{Bhattacharya} S.,  {Kosowsky} A.,  2008, Journal of Cosmology and
  Astro-Particle Physics, 8, 30

\bibitem[\protect\citeauthoryear{{Buchbinder}, {Khoury} \&
  {Ovrut}}{{Buchbinder} et~al.}{2008}]{bko08}
{Buchbinder} E.~I.,  {Khoury} J.,    {Ovrut} B.~A.,  2008, Physical Review
  Letters, 100, 171302

\bibitem[\protect\citeauthoryear{{Carbone}, {Verde} \& {Matarrese}}{{Carbone}
  et~al.}{2008}]{cvm08}
{Carbone} C.,  {Verde} L.,    {Matarrese} S.,  2008, \apjl, 684, L1

\bibitem[\protect\citeauthoryear{{Catelan} \& {Scherrer}}{{Catelan} \&
  {Scherrer}}{1995}]{cs95}
{Catelan} P.,  {Scherrer} R.~J.,  1995, \apj, 445, 1

\bibitem[\protect\citeauthoryear{{Chongchitnan} \& {Silk}}{{Chongchitnan} \&
  {Silk}}{2010}]{chongchitnansilk10}
{Chongchitnan} S.,  {Silk} J.,  2010, ArXiv e-prints, astro-ph/1007.1230

\bibitem[\protect\citeauthoryear{{Cooray} \& {Sheth}}{{Cooray} \&
  {Sheth}}{2002}]{haloreview}
{Cooray} A.,  {Sheth} R.,  2002, \physrep, 372, 1

\bibitem[\protect\citeauthoryear{{Croft} \& {Efstathiou}}{{Croft} \&
  {Efstathiou}}{1995}]{ce94}
{Croft} R.,  {Efstathiou} G.,  1995, in {J.~P.~M{\"u}cket, S.~Gottloeber, \&
  V.~M{\"u}ller} ed., Large Scale Structure in the Universe {Constraints on the
  Power Spectrum of Mass Fluctuations from Galaxy Cluster Peculiar Velocities}.
pp 101--+

\bibitem[\protect\citeauthoryear{{Cunha}, {Huterer} \& {Dore}}{{Cunha}
  et~al.}{2010}]{CHDfnl}
{Cunha} C.,  {Huterer} D.,    {Dore} O.,  2010, ArXiv e-prints,
  astro-ph/1003.2416

\bibitem[\protect\citeauthoryear{{Dalal}, {Dor{\'e}}, {Huterer} \&
  {Shirokov}}{{Dalal} et~al.}{2008}]{dalaletal08}
{Dalal} N.,  {Dor{\'e}} O.,  {Huterer} D.,    {Shirokov} A.,  2008, \prd, 77,
  123514

\bibitem[\protect\citeauthoryear{{Desjacques} \& {Seljak}}{{Desjacques} \&
  {Seljak}}{2010a}]{LSSfnlreview}
{Desjacques} V.,  {Seljak} U.,  2010a, ArXiv e-prints, astro-ph/1003.5020

\bibitem[\protect\citeauthoryear{{Desjacques} \& {Seljak}}{{Desjacques} \&
  {Seljak}}{2010b}]{vincentgnl}
{Desjacques} V.,  {Seljak} U.,  2010b, \prd, 81, 023006

\bibitem[\protect\citeauthoryear{{Desjacques}, {Seljak} \&
  {Iliev}}{{Desjacques} et~al.}{2009}]{fnlvincent}
{Desjacques} V.,  {Seljak} U.,    {Iliev} I.~T.,  2009, \mnras, 396, 85

\bibitem[\protect\citeauthoryear{{Giannantonio} \& {Porciani}}{{Giannantonio}
  \& {Porciani}}{2010}]{gp10}
{Giannantonio} T.,  {Porciani} C.,  2010, \prd, 81, 063530

\bibitem[\protect\citeauthoryear{{Gorski}}{{Gorski}}{1988}]{gorski88}
{Gorski} K.,  1988, \apjl, 332, L7

\bibitem[\protect\citeauthoryear{{Grossi}, {Branchini}, {Dolag}, {Matarrese} \&
  {Moscardini}}{{Grossi} et~al.}{2008}]{grossi08}
{Grossi} M.,  {Branchini} E.,  {Dolag} K.,  {Matarrese} S.,    {Moscardini} L.,
   2008, \mnras, 390, 438

\bibitem[\protect\citeauthoryear{{Grossi}, {Verde}, {Carbone}, {Dolag},
  {Branchini}, {Iannuzzi}, {Matarrese} \& {Moscardini}}{{Grossi}
  et~al.}{2009}]{grossinfm}
{Grossi} M.,  {Verde} L.,  {Carbone} C.,  {Dolag} K.,  {Branchini} E.,
  {Iannuzzi} F.,  {Matarrese} S.,    {Moscardini} L.,  2009, ArXiv e-prints,
  astro-ph/0902.2013

\bibitem[\protect\citeauthoryear{{Hahn}}{{Hahn}}{2005}]{cuba}
{Hahn} T.,  2005, Computer Physics Communications, 168, 78

\bibitem[\protect\citeauthoryear{{Hamana}, {Kayo}, {Yoshida}, {Suto} \&
  {Jing}}{{Hamana} et~al.}{2003}]{hkys03}
{Hamana} T.,  {Kayo} I.,  {Yoshida} N.,  {Suto} Y.,    {Jing} Y.~P.,  2003,
  \mnras, 343, 1312

\bibitem[\protect\citeauthoryear{{Hikage}, {Matsubara}, {Coles}, {Liguori},
  {Hansen} \& {Matarrese}}{{Hikage} et~al.}{2008}]{hikageetal08}
{Hikage} C.,  {Matsubara} T.,  {Coles} P.,  {Liguori} M.,  {Hansen} F.~K.,
  {Matarrese} S.,  2008, \mnras, 389, 1439

\bibitem[\protect\citeauthoryear{{Izumi} \& {Soda}}{{Izumi} \&
  {Soda}}{2007}]{is07}
{Izumi} K.,  {Soda} J.,  2007, \prd, 76, 083517

\bibitem[\protect\citeauthoryear{{Kamionkowski}, {Verde} \&
  {Jimenez}}{{Kamionkowski} et~al.}{2009}]{kvj08}
{Kamionkowski} M.,  {Verde} L.,    {Jimenez} R.,  2009, Journal of Cosmology
  and Astro-Particle Physics, 1, 10

\bibitem[\protect\citeauthoryear{{Khoury} \& {Piazza}}{{Khoury} \&
  {Piazza}}{2008}]{kp08}
{Khoury} J.,  {Piazza} F.,  2008, ArXiv e-prints, hep-th/0811.3633

\bibitem[\protect\citeauthoryear{{Komatsu}}{{Komatsu}}{2010}]{cmbfnlreview}
{Komatsu} E.,  2010, ArXiv e-prints, astro-ph/1003.6097

\bibitem[\protect\citeauthoryear{{Komatsu}, {Smith}, {Dunkley}, {Bennett},
  {Gold}, {Hinshaw}, {Jarosik}, {Larson}, {Nolta}, {Page}, {Spergel},
  {Halpern}, {Hill}, {Kogut}, {Limon}, {Meyer}, {Odegard}, {Tucker}, {Weiland},
  {Wollack} \& {Wright}}{{Komatsu} et~al.}{2010}]{cmb7yr}
{Komatsu} E.,  {Smith} K.~M.,  {Dunkley} J.,  {Bennett} C.~L.,  {Gold} B.,
  {Hinshaw} G.,  {Jarosik} N.,  {Larson} D.,  {Nolta} M.~R.,  {Page} L.,
  {Spergel} D.~N.,  {Halpern} M.,  {Hill} R.~S.,  {Kogut} A.,  {Limon} M.,
  {Meyer} S.~S.,  {Odegard} N.,  {Tucker} G.~S.,  {Weiland} J.~L.,  {Wollack}
  E.,    {Wright} E.~L.,  2010, ArXiv e-prints, astro-ph/1001.4538


\bibitem[\protect\citeauthoryear{{Koyama}, {Soda} \& {Taruya}}{{Koyama}
  et~al.}{1999}]{kst99}
{Koyama} K.,  {Soda} J.,    {Taruya} A.,  1999, \mnras, 310, 1111

\bibitem[\protect\citeauthoryear{{Kuwabara}, {Taruya} \& {Suto}}{{Kuwabara}
  et~al.}{2002}]{kts02}
{Kuwabara} T.,  {Taruya} A.,    {Suto} Y.,  2002, \pasj, 54, 503

\bibitem[\protect\citeauthoryear{{Lam}, {Desjacques} \& {Sheth}}{{Lam}
  et~al.}{2010}]{lsdfnlred}
{Lam} T.~Y.,  {Desjacques} V.,    {Sheth} R.~K.,  2010, \mnras, 402, 2397

\bibitem[\protect\citeauthoryear{{Lam} \& {Sheth}}{{Lam} \&
  {Sheth}}{2009}]{lamshethfnl}
{Lam} T.~Y.,  {Sheth} R.~K.,  2009, \mnras, 395, 1743

\bibitem[\protect\citeauthoryear{{Lam}, {Sheth} \& {Desjacques}}{{Lam}
  et~al.}{2009}]{lsdfnl}
{Lam} T.~Y.,  {Sheth} R.~K.,    {Desjacques} V.,  2009, \mnras, 399, 1482

\bibitem[\protect\citeauthoryear{{Liguori}, {Sefusatti}, {Fergusson} \&
  {Shellard}}{{Liguori} et~al.}{2010}]{cmblssfnlreview}
{Liguori} M.,  {Sefusatti} E.,  {Fergusson} J.~R.,    {Shellard} E.~P.~S.,
  2010, ArXiv e-prints, astro-ph/1001.4707

\bibitem[\protect\citeauthoryear{{Lo Verde}, {Miller}, {Shandera} \&
  {Verde}}{{Lo Verde} et~al.}{2008}]{fnlverde}
{Lo Verde} M.,  {Miller} A.,  {Shandera} S.,    {Verde} L.,  2008, Journal of
  Cosmology and Astro-Particle Physics, 4, 14

\bibitem[\protect\citeauthoryear{{Marian}, {Hilbert}, {Smith}, {Schneider} \&
  {Desjacques}}{{Marian} et~al.}{2010}]{laurafnlWL}
{Marian} L.,  {Hilbert} S.,  {Smith} R.~E.,  {Schneider} P.,    {Desjacques}
  V.,  2010, ArXiv e-prints, arXiv:1010.5242

\bibitem[\protect\citeauthoryear{{Matarrese} \& {Verde}}{{Matarrese} \&
  {Verde}}{2008}]{mv08}
{Matarrese} S.,  {Verde} L.,  2008, \apjl, 677, L77

\bibitem[\protect\citeauthoryear{{Matarrese}, {Verde} \& {Jimenez}}{{Matarrese}
  et~al.}{2000}]{mvj00}
{Matarrese} S.,  {Verde} L.,    {Jimenez} R.,  2000, \apj, 541, 10

\bibitem[\protect\citeauthoryear{{McDonald}}{{McDonald}}{2008}]{mcdonald08}
{McDonald} P.,  2008, \prd, 78, 123519

\bibitem[\protect\citeauthoryear{{McEwen}, {Hobson}, {Lasenby} \&
  {Mortlock}}{{McEwen} et~al.}{2008}]{mhlm08}
{McEwen} J.~D.,  {Hobson} M.~P.,  {Lasenby} A.~N.,    {Mortlock} D.~J.,  2008,
  \mnras, 388, 659

\bibitem[\protect\citeauthoryear{{Nishimichi}, {Taruya}, {Koyama} \&
  {Sabiu}}{{Nishimichi} et~al.}{2009}]{nishimichifnl}
{Nishimichi} T.,  {Taruya} A.,  {Koyama} K.,    {Sabiu} C.,  2009, ArXiv
  e-prints, astro-ph/0911.4768

\bibitem[\protect\citeauthoryear{{Peel}}{{Peel}}{2006}]{peel06}
{Peel} A.~C.,  2006, \mnras, 365, 1191

\bibitem[\protect\citeauthoryear{{Pillepich}, {Porciani} \& {Hahn}}{{Pillepich}
  et~al.}{2008}]{pphfnl08}
{Pillepich} A.,  {Porciani} C.,    {Hahn} O.,  2008, ArXiv e-prints,
  astro-ph/0811.4176

\bibitem[\protect\citeauthoryear{{Rossi}, {Chingangbam} \& {Park}}{{Rossi}
  et~al.}{2010}]{rossicmbfnl}
{Rossi} G.,  {Chingangbam} P.,    {Park} C.,  2010, ArXiv e-prints,
  astro-ph/1003.0272

\bibitem[\protect\citeauthoryear{{Rossi}, {Sheth}, {Park} \&
  {Hernandez-Monteagudo}}{{Rossi} et~al.}{2009}]{rsphmfnl}
{Rossi} G.,  {Sheth} R.~K.,  {Park} C.,    {Hernandez-Monteagudo} C.,  2009,
  ArXiv e-prints, astro-ph/0906.2190

\bibitem[\protect\citeauthoryear{{Sartoris}, {Borgani}, {Fedeli}, {Matarrese},
  {Moscardini}, {Rosati} \& {Weller}}{{Sartoris} et~al.}{2010}]{sbfmmrw10}
{Sartoris} B.,  {Borgani} S.,  {Fedeli} C.,  {Matarrese} S.,  {Moscardini} L.,
  {Rosati} P.,    {Weller} J.,  2010, ArXiv e-prints, astro-ph/1003.0841

\bibitem[\protect\citeauthoryear{{Scherrer}}{{Scherrer}}{1992}]{scherrer92}
{Scherrer} R.~J.,  1992, \apj, 390, 330

\bibitem[\protect\citeauthoryear{{Schmidt}}{{Schmidt}}{2010}]{schmidt10}
{Schmidt} F.,  2010, ArXiv e-prints, astro-ph/1005.4063

\bibitem[\protect\citeauthoryear{{Scoccimarro}}{{Scoccimarro}}{2004}]{rs04}
{Scoccimarro} R.,  2004, \prd, 70, 083007

\bibitem[\protect\citeauthoryear{{Scoccimarro}, {Sefusatti} \&
  {Zaldarriaga}}{{Scoccimarro} et~al.}{2004}]{ssz04}
{Scoccimarro} R.,  {Sefusatti} E.,    {Zaldarriaga} M.,  2004, \prd, 69, 103513

\bibitem[\protect\citeauthoryear{{Sefusatti} \& {Komatsu}}{{Sefusatti} \&
  {Komatsu}}{2007}]{sk07}
{Sefusatti} E.,  {Komatsu} E.,  2007, \prd, 76, 083004

\bibitem[\protect\citeauthoryear{{Seto} \& {Yokoyama}}{{Seto} \&
  {Yokoyama}}{1998}]{seto98}
{Seto} N.,  {Yokoyama} J.,  1998, \apj, 492, 421

\bibitem[\protect\citeauthoryear{{Sheth} \& {Diaferio}}{{Sheth} \&
  {Diaferio}}{2001}]{sdvel01}
{Sheth} R.~K.,  {Diaferio} A.,  2001, \mnras, 322, 901

\bibitem[\protect\citeauthoryear{{Sheth}, {Diaferio}, {Hui} \&
  {Scoccimarro}}{{Sheth} et~al.}{2001}]{sdhs01}
{Sheth} R.~K.,  {Diaferio} A.,  {Hui} L.,    {Scoccimarro} R.,  2001, \mnras,
  326, 463

\bibitem[\protect\citeauthoryear{{Sheth}, {Hui}, {Diaferio} \&
  {Scoccimarro}}{{Sheth} et~al.}{2001}]{shds01}
{Sheth} R.~K.,  {Hui} L.,  {Diaferio} A.,    {Scoccimarro} R.,  2001, \mnras,
  325, 1288

\bibitem[\protect\citeauthoryear{{Sheth} \& {Zehavi}}{{Sheth} \&
  {Zehavi}}{2009}]{sz09}
{Sheth} R.~K.,  {Zehavi} I.,  2009, \mnras, 394, 1459

\bibitem[\protect\citeauthoryear{{Silvestri} \& {Trodden}}{{Silvestri} \&
  {Trodden}}{2008}]{st08}
{Silvestri} A.,  {Trodden} M.,  2008, ArXiv e-prints, astro-ph/0811.2176

\bibitem[\protect\citeauthoryear{{Slosar}}{{Slosar}}{2009}]{slosar08}
{Slosar} A.,  2009, Journal of Cosmology and Astro-Particle Physics, 3, 4

\bibitem[\protect\citeauthoryear{{Slosar}, {Hirata}, {Seljak}, {Ho} \&
  {Padmanabhan}}{{Slosar} et~al.}{2008}]{shshp08}
{Slosar} A.,  {Hirata} C.,  {Seljak} U.,  {Ho} S.,    {Padmanabhan} N.,  2008,
  Journal of Cosmology and Astro-Particle Physics, 8, 31

\bibitem[\protect\citeauthoryear{{Smidt}, {Amblard}, {Byrnes}, {Cooray} \&
  {Munshi}}{{Smidt} et~al.}{2010}]{cmbbktk}
{Smidt} J.,  {Amblard} A.,  {Byrnes} C.~T.,  {Cooray} A.,    {Munshi} D.,
  2010, ArXiv e-prints, astro-ph/1004.1409

\bibitem[\protect\citeauthoryear{{Smith}, {Desjacques} \& {Marian}}{{Smith}
  et~al.}{2010}]{robertfnlhaloModel}
{Smith} R.~E.,  {Desjacques} V.,    {Marian} L.,  2010, ArXiv e-prints,
  arXiv:1009.5085

\bibitem[\protect\citeauthoryear{{Taruya}, {Koyama} \& {Matsubara}}{{Taruya}
  et~al.}{2008}]{tkm08}
{Taruya} A.,  {Koyama} K.,    {Matsubara} T.,  2008, \prd, 78, 123534

\bibitem[\protect\citeauthoryear{{Tseliakhovich}, {Hirata} \&
  {Slosar}}{{Tseliakhovich} et~al.}{2010}]{ths10}
{Tseliakhovich} D.,  {Hirata} C.,    {Slosar} A.,  2010, ArXiv e-prints

\bibitem[\protect\citeauthoryear{{Wands}}{{Wands}}{2010}]{wandslocalfnl}
{Wands} D.,  2010, ArXiv e-prints, astro-ph/1004.0818

\bibitem[\protect\citeauthoryear{{Xia}, {Viel}, {Baccigalupi}, {De Zotti},
  {Matarrese} \& {Verde}}{{Xia} et~al.}{2010}]{nraofnl}
{Xia} J.,  {Viel} M.,  {Baccigalupi} C.,  {De Zotti} G.,  {Matarrese} S.,
  {Verde} L.,  2010, ArXiv e-prints, astro-ph/1003.3451

\bibitem[\protect\citeauthoryear{{Yadav} \& {Wandelt}}{{Yadav} \&
  {Wandelt}}{2008}]{yw08}
{Yadav} A.~P.~S.,  {Wandelt} B.~D.,  2008, Physical Review Letters, 100, 181301

\bibitem[\protect\citeauthoryear{{Yoshida}, {Sheth} \& {Diaferio}}{{Yoshida}
  et~al.}{2001}]{ysd01}
{Yoshida} N.,  {Sheth} R.~K.,    {Diaferio} A.,  2001, \mnras, 328, 669

\end{thebibliography}

\appendix
\section{Derivation of mass weighted linear pairwise velocity PDF}
This appendix derives the expression of the linear
pairwise velocity PDF for $f_{nl}\neq 0$
(equation~\eqref{eqn:qfnl}).  The following derivation also applies to
the derivations of the  linear  velocity difference PDF
when $f_{nl}\neq 0$ (equation~\eqref{eqn:pdffnlvw}, by setting $\delta =
\delta' =0$ in the following derivation) and the 
linear pairwise velocity PDF when $f_{nl}=0$ (equation~\eqref{eqn:q0p0}).

The first step is to generalize the
expression for the line-of-sight pairwise velocity 
generating function $\mathcal{M}$
(equation~(8) in \citet{rs04}) to the 3$d$ generating function of
pairwise velocity:
\begin{equation}
\mathcal{Z}(\bm{\lambda};r) \equiv [1+\xi(r)] \mathcal{M}({\bm
  \lambda};r) = \langle \exp({\bm \lambda \cdot  \bm v})[1+\delta(x)][1+ \delta(x')]\rangle,
\end{equation}
where ${\bm \lambda} \cdot {\bm v} = \lambda\lpa v\lpa 
+ \lambda\lpe[a] v\lpe[a]  + \lambda\lpe[b] v\lpe[b] $ and  $x = x' +r$. The characteristics function can be computed using the
cumulants (see discussion in \citet{rs04}):
\begin{align}
\mathcal{Z}(i{\bm \lambda};r) & = \exp\langle \exp(i{\bm \lambda}\cdot
  {\bm v})\rangle_c[1 + \lrangle{\exp(i {\bm \lambda}\cdot {\bm v})
    (\delta + \delta')}_c + \lrangle{\exp(i{\bm \lambda} \cdot {\bm
      v})\delta}_c\lrangle{\exp(i{\bm \lambda} \cdot {\bm
      v})\delta}_c + \lrangle{\exp(i{\bm \lambda} \cdot {\bm
      v})\delta\delta'}_c] \\
& =
\exp\left[\sum_j \frac{\lrangle{v_j^2}}{2}(i\lambda_j)^2 +
  \frac{\lrangle{v\lpa ^3}}{3!}(i\lambda\lpa )^3
  +\frac{\lrangle{v\lpa v\lpe[a] ^2}}{2}(i\lambda\lpa )(i\lambda\lpe[a] )^2
  +
  \frac{\lrangle{v\lpa v\lpe[b] ^2}}{2}(i\lambda\lpa )(i\lambda\lpe[b] )^2
+ \dots \right] \nonumber \\
& \qquad \times \left\{1  + \xi(r)  +
                            \lrangle{v\lpa \delta\delta'}(i\lambda\lpa )
                            + \sum_j \left[\lrangle{v_j(\delta+\delta')}(i\lambda_j) + 
                        \frac{\lrangle{v_j^2(\delta
                            +\delta')}}{2}(i\lambda_j)^2 + \dots
                      \right] \right. \nonumber \\
 & \qquad \quad \left.   + \left[ \sum_j \lrangle{v_j\delta}(i\lambda_j) + 
                              \frac{\lrangle{v_j^2\delta}}{2}(i\lambda_j)^2
                            + \dots \right] 
\left[ \sum_j \lrangle{v_j\delta'}(i\lambda_j) + 
                              \frac{\lrangle{v_j^2\delta'}}{2}(i\lambda_j)^2
                            + \dots \right]
                                            + \dots\right\},
\end{align}
where $\delta \equiv \delta(x)$, $\delta' \equiv \delta(x')$, and the
summation over $j$ denotes  summing over all three directions 
$\{\parallel,\perp_a,\perp_b\}$. 
The first term in the second equity is the
linear  velocity difference characteristic function 
while the second term is associated with the mass weighting.
The characteristic function for the linear
velocity difference for $f_{nl}=0$,
\begin{equation}
\mathcal{F}_0(i{\bm \lambda}; r) = \exp\left[\sum_j
  \frac{\lrangle{v_j^2}}{2}(i\lambda_j)^2 \right],
\end{equation}
whose Fourier transofrm is the  linear 
velocity difference PDF when $f_{nl}=0$ (equation~\eqref{eqn:pdfgvw}).
Hence we can factorize $\mathcal{F}_0$ from $\mathcal{Z}$ and define 
$\mathcal{G}$
\begin{equation}
\mathcal{Z}(i{\bm \lambda};r ) \equiv  \mathcal{G}(i{\bm \lambda};r
)\mathcal{F}_0(i{\bm \lambda};r) 
\end{equation}

So far the expression is exact. 
Now we take the approximation that only terms up to the first order
in $f_{nl}$ is included and notice that 
$\lrangle{v\lpe[a] \delta} = \lrangle{v\lpe[b] \delta} = 0$. Hence
\begin{align}
\mathcal{G}(i{\bm \lambda};r)& = 1 + \xi(r) +\left(\lrangle{v\lpa \delta} +
\lrangle{v\lpa \delta'} + \lrangle{v\lpa\delta\delta'} \right)(i\lambda\lpa)
+ \lrangle{v\lpa\delta}\lrangle{v\lpa\delta'}(i\lambda\lpa)^2
+ \left[\frac{\lrangle{v\lpa^3}}{3!}(1+\xi(r)) +
  \frac{\lrangle{v\lpa\delta}\lrangle{v\lpa^2\delta'}}{2} +
  \frac{\lrangle{v\lpa\delta'}\lrangle{v\lpa^2\delta}}{2}\right](i\lambda\lpa)^3
\nonumber \\ 
&
 + \frac{\lrangle{v\lpa^3}}{3!}\left(\lrangle{v\lpa\delta} +
   \lrangle{v\lpa\delta'}\right)(i\lambda\lpa)^4
+
\frac{\lrangle{v\lpa^3}}{3!}\lrangle{v\lpa\delta}\lrangle{v\lpa\delta'}(i\lambda\lpa)^5
 + \sum_{\alpha=a,b}\Gamma(i\lambda\lpe[\alpha]),
\end{align}
where 
\begin{align}
\Gamma(i \lambda\lpe[\alpha])
& = 
\left[\frac{\lrangle{v\lpa\delta}\lrangle{v\lpe[\alpha]^2\delta'}}{2} + 
               \frac{\lrangle{v\lpa\delta'}\lrangle{v\lpe[\alpha]^2\delta}}{2}
    +\frac{\lrangle{v\lpa
        v\lpe[\alpha]^2}}{2}(1+\xi(r))\right](i\lambda\lpa)(i\lambda\lpe[\alpha])^2+\frac{\lrangle{v\lpa
      v\lpe[\alpha]^2}}{2}\left(\lrangle{v\lpa\delta} +
    \lrangle{v\lpa\delta'}\right)(i\lambda\lpa)^2(i\lambda\lpe[\alpha])^2
  \nonumber \\
& + \frac{\lrangle{v\lpa v\lpe[\alpha]^2}}{2}\lrangle{v\lpa\delta}\lrangle{v\lpa\delta'}(i\lambda\lpa)^3(i\lambda\lpe[\alpha])^2.
\end{align}

To get the mass weighted linear PDF we need to do a Fourier transform
on $\mathcal{Z}$. While the expression seems very complicated, the
Fourier transform can be done analytically due to the fact that the
Fourier transform of 
$\mathcal{G}$ is actually a derivative operator acting on the uniform
weighted linear PDF. Recall that the uniform weighted linear PDF is a
product of three univariate PDFs, hence the derivative
operation can be done easily and the resulting mass weighted linear
pairwise velocity PDF is equation~\eqref{eqn:qfnl}.

\label{lastpage}
\end{document}